\documentclass[11pt]{article}

\usepackage[final]{acl}

\usepackage{times}
\usepackage{latexsym}

\usepackage[T1]{fontenc}

\usepackage[utf8]{inputenc}

\usepackage{microtype}

\usepackage{inconsolata}

\usepackage{graphicx}

\usepackage{enumitem}
\usepackage{amsmath}
\usepackage{comment}
\usepackage{hyperref}
\usepackage{amsmath}
\usepackage{mathtools}
\usepackage{amsfonts}
\usepackage{listings}
\usepackage{amssymb}
\usepackage{makecell}

\title{Training LLMs to Enforce Multi-Level Instruction Hierarchies via Gravity-Weighted Direct Preference Optimization}

\author{Lena S. Bolliger,~\;~Lena A. Jäger\\
Department of Computational Linguistics, University of Zurich, Switzerland\\
\texttt{\{\href{mailto:bolliger@cl.uzh.ch}{bolliger},\href{mailto:jaeger@cl.uzh.ch}{jaeger}\}@cl.uzh.ch}}

\begin{document}
\maketitle

\begin{abstract}
Production LLMs receive instructions from sources with very different levels of trust, yet attend to every token with uniform architectural privilege. This is the structural vulnerability that enables malicious prompt injections and, more broadly, leaves models without a principled way to resolve conflicts between legitimate but competing instructions. A common training-based response is to teach models an explicit instruction hierarchy; existing approaches, however, formalize hierarchies of only three or four levels, treat all violations as equally severe, and rarely evaluate the full set of pairwise level interactions. We formalize a $k$-level instruction hierarchy problem and instantiate it for $k=5$, yielding ten pairwise priority relations that a compliant model must enforce. We then introduce Gravity-Weighted DPO (GW-DPO), a preference-optimization objective whose per-sample offset scales with the structural distance between conflicting levels under a linear or bilateral schedule, the latter weighting severity by both the privilege gap and the privilege of the victim level. Combined with hierarchy-specific delimiter tokens~\citep{chen2025struq} and Instructional Segment Embeddings~\citep[ISE]{wu2024instructional}, GW-DPO with the bilateral schedule Pareto-improves over standard DPO and the linear variant on Llama-3.1-8B-Instruct, raising macro pairwise priority adherence while keeping over-refusal at half the standard DPO rate. Ablations isolate ISE as a refusal-threshold calibrator and recast five- versus three-level training as a generality--specialization tradeoff.
\end{abstract}

\section{Introduction}
\label{sec:introduction}

Large language models (LLMs) deployed in production handle instructions from sources that have very different levels of trust. A platform provider sets immutable safety policies. An application developer writes a system prompt that defines persona, scope, and behavior. An end-user types a query. A retrieval pipeline or tool returns content that may have been written by an adversary. All of this gets concatenated into a single token sequence, and the model attends to every token with the same architectural privilege. \citet{wallace2024instruction} describe this as execution ``in kernel mode'', where every input carries maximum authority regardless of where it comes from. 

This uniformity is the structural cause of \textit{prompt injection}: adversarial instructions hidden in inputs that hijack the model's intended behavior~\citep{willison2022prompt}. The common diagnosis is that LLMs lack any mechanism to distinguish instructions from data, or to enforce a priority ordering when instructions from different sources conflict --- a vulnerability now documented across nearly every deployment surface (cf.~\S~\ref{sec:related-work}).

\citet{wallace2024instruction} proposed the first training-based response: a three-level instruction hierarchy with synthetic conflict data and SFT+RLHF on GPT-3.5~\citep{brown2020language}. A line of follow-up work extends this with delimiter tokens, task-specific fine-tuning, preference optimization, and learnable per-token segment signals (cf.~\S~\ref{sec:related-work}), but no published recipe trains a hierarchy beyond three or four levels, encodes the severity of a violation in the loss, or evaluates against the full set of pairwise level interactions.

The gap between the literature and what production systems already require is wider than three levels: OpenAI's 2025 Model Spec names five stakeholder tiers and Anthropic's Usage Policy and Constitutional Principles operate as a similar multi-tier stack. Classical computer security has known how to think about hierarchical privilege for fifty years: Biba's integrity model~\citep{biba1977integrity} formalized that low-integrity inputs should not ``write up'' into high-integrity behavior, and \citet{denning1976lattice} showed that the rules governing information flow can be derived directly from the lattice structure of the privilege labels. The mapping to LLMs is direct: an instruction hierarchy is a lattice, the rule that tool outputs must not override platform safety is a ``no write up'' constraint, and violations at the top are inherently more severe than violations at the bottom. We instantiate it for $k=5$ levels.

This paper makes three contributions. First, we formalize the $k$-level instruction hierarchy problem and instantiate it for $k=5$ with a concrete level structure 
grounded in the production architectures cited above. The formalization produces ten pairwise priority relationships that any compliant model must enforce, and we construct a custom evaluation suite that tests all ten. 
Second, we introduce Gravity-Weighted DPO (GW-DPO), a preference optimization objective that adapts ODPO~\citep{amini-etal-2024-direct} by setting the per-sample margin proportional to the hierarchy distance between the conflicting levels, and we evaluate two specific schedules: a \textit{linear}  
and a \textit{bilateral} form, 
the latter of which scales severity by both the privilege gap and the privilege of the victim level. The bilateral schedule is the closer realization of the gravitational analogy in \textit{Gravity-Weighted} DPO, and it Pareto-improves over both standard DPO and the linear schedule on conflict-resolution accuracy at a fraction of the over-refusal cost. 
Third, we combine the loss with two architectural modifications introduced by previous research (Instructional Segment Embeddings (ISE,~\citealp{wu2024instructional}) and special delimiter tokens, as suggested in \citet{chen2025struq}) and isolate the contribution of each component through a suite of ablations. We show that ISE primarily calibrates the model's refusal threshold rather than its raw conflict-resolution accuracy, and that five- versus three-level training emerges as a generality--specialization tradeoff rather than a simple ordering question. 
The code to reproduce the data generation and model training pipeline is available via a \href{https://github.com/DiLi-Lab/gw-dpo-prompt-hierarchy}{GitHub repository}.

\section{Related Work}
\label{sec:related-work}

\paragraph{Prompt injection and the instruction-data boundary.} \citet{greshake2023} demonstrated indirect prompt injection through retrieved content, and \citet{perez2022ignore} systematized the direct variant. The vulnerability is now documented across multi-turn dialogue~\citep{russinovich2025great}, retrieval-augmented memory~\citep{zou2023universal, chen2024agentpoison}, tool-using agents~\citep{debenedetti2024agentdojo, zhan-etal-2024-injecagent}, and adaptive attacks designed to defeat training-based defenses~\citep{nasr2025attacker}. The benchmarking landscape has begun to formalize the underlying instruction-conflict problem: \citet{zhang-etal-2025-iheval} construct a multi-axis hierarchy benchmark on which current open-source LLMs score roughly half as well on conflicts as on aligned controls, and \citet{geng2025control} show that deployed hierarchies can be overridden by authority framings such as claimed expertise.

\paragraph{Instruction hierarchies and architectural separation.} Two complementary lines have emerged from \citet{wallace2024instruction}'s original three-level recipe. On the training side, \citet{chen2025struq}, \citet{piet2024jatmo}, and \citet{chen2025secalign} train models to separate instructions from data using delimiters, task-specific fine-tuning, or preference optimization; SecAlign~\citep{chen2025secalign} in particular reports a $2.8\times$ larger log-probability margin between compliant and non-compliant responses under DPO than under SFT alone, the empirical basis for our choice of DPO as the primary hierarchy-training objective. On the architectural side, \citet{wu2024instructional} introduced Instructional Segment Embeddings (ISE), a learnable per-token signal indexed by hierarchy level, and \citet{zverev2025aside} applied orthogonal rotations to data tokens to encode instruction-vs-data separation. We adopt elements from both lines (SecAlign's preference framing, StruQ's delimiter tokens, ISE) and extend the hierarchy beyond the three- or four-level ceiling these works share.

\paragraph{Preference optimization and the gravity-weighted objective.} Direct Preference Optimization (DPO,~\citealp{rafailov2023direct}) replaces the explicit reward model of RLHF with a closed-form loss derived from the Bradley-Terry preference model~\citep{bradley1952rank}. Three refinements of DPO are directly relevant to the current study: \citet{amini-etal-2024-direct} introduced ODPO, which adds a per-sample offset $\delta$ that imposes a minimum reward gap; \citet{pattnaik-etal-2024-enhancing} demonstrated in Curry-DPO that training on large-margin preference pairs before small-margin ones improves performance; and \citet{kim-etal-2025-sdpo} showed in sDPO that updating the reference model between training stages prevents excessive divergence.

\section{Problem Setting}
\label{sec:problem-setting}

We define a $k$-level instruction hierarchy $H = (L_0, L_1, \dots, L_{k-1})$ for language models, where $L_0$ is the highest-privilege level and $L_{k-1}$ is the lowest. For any pair of levels $L_i$ and $L_j$, where $i < j$, the model must satisfy the following priority constraint: if the instruction at $L_j$ conflicts with the instruction at $L_i$, the model must follow $L_i$. If they are compatible, the model follows both. For $k=5$, we instantiate: $L_0$: Platform Governance; $L_1$: Developer System Prompt; $L_2$: Per-User Configuration; $L_3$: User Message; $L_4$: Data/Tool Outputs.\footnote{For a conceptual description of the instruction hierarchy levels, please refer to Table~\ref{tab:instruction-levels-description} in Appendix~\ref{appendix:methodology}.} 
This produces 10 pairwise priority relationships that the model must enforce: ($L_0$--$L_1$), ($L_0$--$L_2$), ($L_0$--$L_3$), ($L_0$--$L_4$), ($L_1$--$L_2$), ($L_1$--$L_3$), ($L_1$--$L_4$), ($L_2$--$L_3$), ($L_2$--$L_4$), ($L_3$--$L_4$).  

\section{5-Level Hierarchy Dataset Construction}
\label{sec:dataset-construction}

We construct three datasets purpose-built for the five-level setting: an SFT dataset, a DPO dataset, and an evaluation suite. All three are built as disjoint splits over a common pool of base datasets and content libraries.

\subsection{Base Datasets}
\label{sec:base-datasets} 

We use two instruction-following datasets as raw material: Alpaca Cleaned\footnote{\url{https://huggingface.co/datasets/yahma/alpaca-cleaned}}, which is also the training source for SecAlign~\citep{chen2025secalign} and StruQ~\citep{chen2025struq}, and Dolly 15K\footnote{\url{https://huggingface.co/datasets/databricks/databricks-dolly-15k}}, whose longer \texttt{context} passages and richer task-category metadata support $L_4$ population and $L_1$ persona matching. Field-to-level mappings, partitioning, and spurious-correlation safeguards are detailed in Appendix~\ref{appendix:base-datasets}.

\subsection{Content Libraries}
\label{sec:content-libraries}
Each hierarchy level is filled from a dedicated content library, built by a shared three-step recipe: extraction from published source documents (policies, style guides, task taxonomies), LLM-assisted expansion with Claude Sonnet 4, and deduplication by sentence-embedding cosine similarity. Two auxiliary libraries support adversarial training: a hand-crafted set of injection templates and a Claude-generated library of $L_0$ conflict scenarios for the most subtle $L_0$-involving DPO conflict pairs. Per-level schemas, source documents, and validation steps are detailed in Appendix~\ref{appendix:content-libraries}.

\subsection{SFT Dataset}
\label{sec:sft-dataset}

$\mathcal{D}_{\text{SFT}}$ (\verb|~|10k train / \verb|~|1.6k validation) teaches the model to recognize the new delimiter tokens and integrate ISE information. Each instance is a 5-level prompt wrapped in delimiters (cf.~\S~\ref{sec:architecture}), drawn from three categories: \textit{aligned} (\verb|~|7k, including 2k built via context synthesis~\citep{wallace2024instruction}) covers conflict-free 5-level prompts; \textit{partial level} (\verb|~|2k) populates only some tiers so the model learns that not every prompt carries all five; and \textit{trivially misaligned} (\verb|~|1k) seeds a basic refusal signal across four obvious conflict pairs ($L_0$--$L_3$, $L_1$--$L_3$, $L_1$--$L_4$, $L_3$--$L_4$). Assembly path, response-grounded $L_2$ generation, and context synthesis are detailed in Appendix~\ref{appendix:sft-dataset}.

\subsection{DPO Dataset}
\label{sec:dpo-dataset}

$\mathcal{D}_{\text{DPO}}$ (\verb|~|8.7k train / \verb|~|2.5k validation) teaches hierarchy compliance. Each example is a triple $(x, y_w, y_l)$, where $x$ is a 5-level prompt containing a conflict between levels $L_i$ and $L_j$ ($i<j$), $y_w$ correctly follows $L_i$, and $y_l$ incorrectly follows $L_j$. We adapt SecAlign's Algorithm 1~\citep{chen2025secalign} to all ten pair types: seven use base-dataset concatenation for $y_l$, the three $L_0$-involving pairs that require nuanced refusals ($L_0$--$L_1$, $L_0$--$L_2$, $L_0$--$L_4$) use context distillation~\citep{wallace2024instruction} via Claude Sonnet~4 for $y_w$ and GPT-4o-mini for $y_l$, and the $L_0$--$L_3$ pair draws $y_w$ from a 15-variant refusal template pool. To prevent over-refusal we add 2k benign-looking calibration pairs, plus 1k cascading examples with 3+ simultaneous conflicts. All pairs pass a two-stage quality-control pipeline (automated filters followed by independent dual-judge scoring by GPT-4o and Gemini~2.5~Pro); per-pair construction, the cascading families, the post-phase repair pipeline, and the full QC protocol are in Appendix~\ref{appendix:dpo-dataset} and Appendix~\ref{appendix:quality-control}.

\subsection{Evaluation Suite}
\label{sec:evaluation-suite}

Since no existing benchmark tests all ten pairwise conflict relationships across five hierarchy levels, we construct a dedicated evaluation suite of \verb|~|1750 instances split across three roles. The \textit{conflict} split tests whether the model resolves a deliberate conflict between two levels in favor of the higher-privileged one. The \textit{aligned controls} split replaces the attacking content with compatible content, keeping everything else fixed; the same model that resolves a conflict correctly should not refuse the matched benign prompt. The \textit{reference baselines} split presents the same conflict scenarios as flat plain-text concatenations with all delimiters stripped, isolating how much performance is attributable to the architectural cues versus the underlying instruction-following capability. 
Scenarios are composed holistically by GPT-4o (rather than independent per-level assembly), with gold responses by Claude Sonnet~4 and strict dual-judge filtering. Cross-split exclusion guarantees that no base-dataset row used in SFT or DPO appears in the evaluation suite, making this, to our knowledge, the first benchmark to cover all ten pairwise level interactions at $k=5$. Pipeline, material selection, and QC thresholds are detailed in Appendix~\ref{appendix:evaluation-suite}.

\section{Gravity-Weighted DPO}
\label{sec:gravity-weighted-DPO}

The DPO training phase (cf.~\S~\ref{sec:training-dpo}) teaches the model to comply with the instruction hierarchy through contrastive preference optimization. Standard DPO~\citep{rafailov2023direct} treats every preference pair identically and thus every hierarchy violation as equally severe, while a platform-policy breach induced by an injected tool output is structurally more dangerous than a user request being overridden by retrieved content. We adapt the offset variant of DPO (ODPO,~\citealp{amini-etal-2024-direct}), which adds a per-sample, non-negative offset $\delta\geq 0$ inside the sigmoid. ODPO parameterizes $\delta$ as a monotone function of a continuous quality gap between $y_w$ and $y_l$ (\emph{e.g.,} $\delta=\alpha\cdot\log(r(y_w)-r(y_l))$ when scalar reward scores are available) and proves that this enforces a target log-ratio gap $\Delta r_\theta\geq\delta/\beta$ at the optimum. The ODPO design leaves $\delta$ open, and we exploit this freedom to encode a \textit{structural} rather than a \textit{quality-based} preference signal: we annotate each preference triple $(x,y_w,y_l)$ in $\mathcal{D}_{\text{DPO}}$ with a \textit{victim level} $i\in\{0,\dots,k-1\}$, whose instruction $y_w$ correctly follows, and an \textit{attacker level} $j\in\{0,\dots,k-1\}$ with $i\leq j$, whose instruction $y_l$ incorrectly follows.\footnote{Calibration pairs (cf.~\S~\ref{sec:dpo-dataset}) carry $i=j$ and represent benign lower-level instructions that should be obeyed.} The Gravity-Weighted DPO (GW-DPO) loss is then
\begin{equation}
\begin{aligned}
\mathcal{L}_{\text{GW-DPO}}&(\pi_\theta;\pi_{\text{ref}}) \\
= -\mathbb{E}_{(x,y_w,y_l,i,j)\sim\mathcal{D}_{\text{DPO}}}\Big[ &\log\sigma\big(
\underbrace{\beta\cdot\Delta r_\theta(x,y_w,y_l)}_{\text{standard DPO}} \\
&-\,\underbrace{\delta(i,j)}_{\substack{\text{ODPO offset,}\\ \text{level-pair indexed}}}
\big)\Big],
\end{aligned}
\end{equation}
where the implicit reward margin is
\begin{equation}
\small
\begin{aligned}
\Delta r_\theta(x,y_w,y_l)
&= \log\frac{\pi_\theta(y_w\mid x)}{\pi_{\text{ref}}(y_w\mid x)} \\
&\quad - \log\frac{\pi_\theta(y_l\mid x)}{\pi_{\text{ref}}(y_l\mid x)}.
\end{aligned}
\end{equation}
Here $\pi_\theta$ is the policy being trained, $\pi_{\text{ref}}$ the frozen post-SFT reference checkpoint, $\sigma(\cdot)$ the logistic sigmoid, and $\beta>0$ the inverse-temperature parameter. The \textit{margin schedule} $\delta:\{0,\dots,k-1\}^2\to\mathbb{R}_{\geq 0}$ satisfies $\delta(i,i)=0$ for every $i$ and $\delta(i,j)>0$ whenever $i<j$: calibration pairs collapse the loss to standard DPO 
while conflicts incur a strictly positive margin. The ODPO reward-gap bound carries over in pair-specific form, $\Delta r_\theta(x,y_w,y_l)\geq\delta(i,j)/\beta$ (cf. Appendix~\ref{appendix:inherited-reward-gap-bound}). Standard DPO and ODPO background is in Appendix~\ref{appendix:dpo-odpo-background}. We instantiate two concrete schedules.

\paragraph{Bilateral schedule.} The bilateral schedule factorizes severity into the privilege gap and the privilege of the victim level:
\begin{equation}
\small
    \delta_{\text{bi}}(i,j)\;=\;\alpha\cdot(j-i)\cdot(k-1-i), 
\end{equation}
where $\alpha > 0$ is a scaling coefficient and $(k-1-i)$ counts the strictly less privileged levels below $L_i$ (0 only at the lowest level, which cannot be a victim). The first factor preserves the gap-aware signal that distant conflicts are clearer than adjacent ones; the second amplifies the loss whenever a high-privilege level is the victim, regardless of attacker distance. 

\paragraph{Linear schedule.} The linear schedule depends only on the privilege gap:
\begin{equation}
\small
    \delta_{\text{lin}}(i,j)\;=\;\alpha\cdot(j-i).
\end{equation}
It treats every gap-one conflict ($L_0$-vs-$L_1$ as much as $L_3$-vs-$L_4$) as equally severe, which is exactly the assumption the bilateral schedule relaxes. Per-pair offsets for both schedules are 
in Appendix~\ref{appendix:per-pair-offsets}.

\section{Experiments}
\label{sec:experiments}

\subsection{Architecture}
\label{sec:architecture}

We use Llama-3.1-8B-Instruct~\citep{grattafiori2024llama3herdmodels} as the base model 
and modify it with two complementary mechanisms:

\textbf{\textit{Special delimiter tokens.}} Following StruQ~\citep{chen2025struq}, we introduce 12 new special tokens to the tokenizer vocabulary, which are start and end delimiters for each of the five hierarchy levels: \texttt{<|L0\_START|>, <|L0\_END|>, ..., <|L4\_START|>, <|L4\_END|>}, and additional response delimiters \texttt{<|RESP\_START|>, <|RESP\_END|>}. We initialize them from the mean of all existing token embeddings.

\textbf{\textit{Instructional Segment Embeddings (ISE).}} Following \citet{wu2024instructional}, we add a learnable embedding layer: 
each token receives an additive segment embedding based on which hierarchy level it belongs to. The final embedding $E_f\left[\cdot\right]$ for an input token $x_t$ is defined as $E_f\left[x_t\right]$ = $E_{tok}\left[x_t\right]$ + $E_{seg}\left[h_t\right]$, where $E_{tok}$ is the token embedding, $E_{seg}$ the segment embedding, and $h_t \in \left\{0, 1, 2, 3, 4, 5 \right\}$ is determined by parsing the delimiter structure. 
The ISE layer is initialized from $\mathcal{N}(0, 0.0001)$. 
Since Llama's RoPE~\citep{su2023roformerenhancedtransformerrotary} is applied within the attention blocks, ISE does not interfere with the position information.

\subsection{Training}
\label{sec:training}

\subsubsection{Phase 1: Minimal-Adapter SFT}
\label{sec:training-sft}

We fine-tune Llama-3.1-8B-Instruct on $\mathcal{D}_{\text{SFT}}$ (\S\ref{sec:sft-dataset}) with LoRA adapters on the attention and gated-MLP projections. The 12 new delimiter token embeddings and the ISE embedding layer are trained alongside, and the remainder of the base model is frozen. Phase~1 establishes delimiter and ISE recognition; hierarchy compliance is deferred to Phase~2. Full hyperparameters in Appendix~\ref{appendix:training-hyperparameters}.

\subsubsection{Phase 2: GW-DPO}
\label{sec:training-dpo}

We initialize the policy $\pi_\theta$ from the best Phase~1 checkpoint with its LoRA adapters merged into the base, and freeze a copy of the same checkpoint as the reference $\pi_{\text{ref}}$. We optimize the GW-DPO loss with the bilateral or linear schedule on $\mathcal{D}_{\text{DPO}}$ (\S\ref{sec:dpo-dataset}), using LoRA on the same projections as Phase~1; the 12 delimiter token embeddings and the ISE layer remain trainable. Training follows a three-stage easy-to-hard curriculum on cumulative subsets of $\mathcal{D}_{\text{DPO}}$ (\S\ref{sec:curriculum-learning}). The DPO inverse temperature $\beta=0.1$ and gravity coefficient $\alpha=0.3$ come from the sweep in \S\ref{sec:hyperparameter-tuning}; remaining hyperparameters are in Appendix~\ref{appendix:training-hyperparameters}.

\paragraph{Curriculum Learning}
\label{sec:curriculum-learning}
The GW-DPO margin shrinks with hierarchy gap, making small-gap conflicts both weakly supervised and hence harder to learn. Following \citet{pattnaik-etal-2024-enhancing}, we use an easy-to-hard curriculum over the gap $j-i$ with cumulative filtering: stage 1 trains on gap $\geq 3$, stage 2 adds gap-2 pairs, and stage 3 includes adjacent conflicts. Calibration pairs ($i=j$) appear in all stages to prevent refusal bias. Following \citet{kim-etal-2025-sdpo}, we update both the reference and the policy between stages using the best-of-stage checkpoint. Details on the curriculum staging and reference-model updates are in Appendix~\ref{appendix:curriculum-reference-model}.

\subsection{Hyperparameter Tuning}
\label{sec:hyperparameter-tuning}

GW-DPO utilizes, in addition to the inverse temperature $\beta$, also the gravity coefficient $\alpha$. Because the implicit-reward target gap satisfies $\Delta r_\theta\geq\delta(i,j)/\beta$, the two interact and $\alpha$ is meaningful only relative to $\beta$. We therefore reparametrize the search in the ratio $\rho = \alpha / \beta$ 
and tune $(\rho, \beta)$ directly, deriving $\alpha=\rho \cdot \beta$. We score each configuration by its \textit{macro-averaged reward accuracy} on a held-out, stratified cut of the validation set of $\mathcal{D}_{\text{DPO}}$. Reward accuracy depends only on the order of the implicit rewards and is thus scale-invariant in $\beta$. The selected configuration is $\rho=3$, $\beta=0.1$, $\alpha=0.3$. Appendix~\ref{appendix:hyperparameter-tuning} reports the search grid, held-out dataset construction, per-configuration training flow, per-phase results, and limitations.


\subsection{Reference and Ablation Studies}
\label{sec:reference-ablations}

Beyond GW-DPO~(bi.) and GW-DPO~(lin.), we evaluate five configurations that isolate individual system components. Two reference floors bound the table from below: \textit{baseline}, off-the-shelf Llama-3.1-8B-Instruct with only the 12 delimiter tokens added to its tokenizer, and \textit{SFT}, the Phase~1 checkpoint without DPO. \textit{DPO} sets $\alpha=0$ so the GW-DPO loss reduces to standard DPO; the DPO$\rightarrow$GW-DPO~(lin.)$\rightarrow$GW-DPO~(bi.) progression is our headline comparison for the structural-margin claim. \textit{No-ISE} is GW-DPO~(lin.) without the ISE layer. \textit{3-level~(lin.)} collapses $L_0$, $L_1$, $L_2$ into a single \texttt{System} tier in the hierarchy style of \citet{wallace2024instruction}, with all other hyperparameters held at GW-DPO~(lin.) values; we refer to the collapsed levels as \textit{intra-System}. The GW-DPO~(lin.)$\rightarrow$3-level~(lin.) gap attributes to the depth of the training hierarchy and grounds the five-vs-three level verdict in \S\ref{sec:discussion}. Full configurations, design choices, and curriculum adaptations for the 3-level run are in Appendix~\ref{appendix:reference-ablation-studies}.

\subsection{Evaluation Results}
\label{sec:experiments-evaluation-results}

\subsubsection{Evaluation Metrics}
\label{sec:evaluation-metrics}

We report three families of metrics, each computed on a different split of the evaluation suite (\S\ref{sec:evaluation-suite}). \textit{Hierarchy resolution} (conflict split) asks whether the model resolves a conflict in favor of the higher-privileged level: per pair, the Pairwise Priority Adherence ($\mathrm{PPA}_{i,j}$) is the fraction of conflict scenarios on which the model satisfies $L_i$ without also following $L_j$; we aggregate it as a macro mean over all ten pairs ($\mathrm{PPA}_{\mathrm{macro}}$, matching the equal-weight convention of \citealp{wallace2024instruction} and \citealp{zhang-etal-2025-iheval}) and as our own gap-weighted Weighted Hierarchy Score ($\mathrm{WHS}$) that scales each pair's contribution by the privilege gap $j-i$, analogous to the gravity-weighted training objective. The complementary Attack Success Rate $\mathrm{ASR}_{i,j}=1-\mathrm{PPA}_{i,j}$ is reported for comparability with the prompt-injection literature~\citep{toyer2024tensor}. \textit{Generalization} (reference split) asks whether hierarchy reasoning transfers to flat-text rewrites of the conflict scenarios that strip all level wrappers: we apply the same PPA aggregates, denoted $\mathrm{PPA}^{\mathrm{ref}}$, and additionally report the per-pair Utility Delta $\Delta_{i,j}=\mathrm{PPA}_{i,j}-\mathrm{PPA}^{\mathrm{ref}}_{i,j}$ with its signed and absolute means, following IHEval~\citep{zhang-etal-2025-iheval}. \textit{Safety calibration} (aligned-control split) asks whether the model wrongly refuses benign prompts that superficially resemble adversarial ones: the Over-Refusal Rate ($\mathrm{ORR}$) is the fraction of paired benign prompts on which the model refuses, in the spirit of the false-refusal axis of XSTest~\citep{roettger2024xstesttestsuiteidentifying}. 
Full definitions and aggregation rules are in Appendix~\ref{appendix:evaluation-metrics}.

\subsubsection{5-Level Evaluation Suite}
\label{sec:5-level-evaluation-suite}

Table~\ref{tab:eval-headline} reports the headline aggregates over the \textit{conflict}, \textit{reference}, and \textit{aligned-control} instances of the suite (\S\ref{sec:evaluation-suite}). All judge-derived fields are produced by GPT-4o; per-pair tables and additional diagnostics are deferred to Appendix~\ref{appendix:eval-results}.

\begin{table*}[t]
\centering
\small
\begin{tabular}{l c c c c c c c}
\hline
Metric & baseline & SFT & GW-DPO (bi.) & GW-DPO (lin.) & DPO & No-ISE & 3-level~(lin.) \\
\hline
$\mathrm{PPA}_{\mathrm{macro}}$ ($\uparrow$)        & 0.245 & 0.332 & \textbf{0.838} & 0.719 & 0.814 & 0.677 & 0.778 \\
$\mathrm{WHS}$ ($\uparrow$)                          & 0.271 & 0.388 & \textbf{0.885} & 0.751 & 0.852 & 0.728 & 0.826 \\
$\mathrm{PPA}^{\mathrm{ref}}_{\mathrm{macro}}$ ($\uparrow$) & 0.660 & 0.493 & 0.793 & \textbf{0.843} & 0.790 & 0.810 & 0.780 \\
$\bar{\Delta}$ ($\to 0$)                             & $-0.415$ & $-0.161$ & $+0.044$ & $-0.125$ & $\mathbf{+0.024}$ & $-0.133$ & $-0.002$ \\
$\bar{\Delta}_{\mathrm{abs}}$ ($\downarrow$)         & 0.415 & 0.214 & 0.084 & 0.143 & \textbf{0.053} & 0.149 & 0.084 \\
$\mathrm{ORR}$ ($\downarrow$)                        & 0.044 & 0.092 & 0.057 & \textbf{0.024} & 0.120 & 0.576 & 0.074 \\
Avg.\ conflict score, 1--5 ($\uparrow$)              & 2.15  & 2.42  & \textbf{4.41} & 3.88 & 4.30  & 3.75  & 4.26  \\
Conflict completion ($\uparrow$)                     & 0.491 & 0.483 & \textbf{1.000} & \textbf{1.000} & \textbf{1.000} & \textbf{1.000} & \textbf{1.000} \\
Aligned completion ($\uparrow$)                      & 0.563 & 0.524 & \textbf{1.000} & \textbf{1.000} & \textbf{1.000} & \textbf{1.000} & \textbf{1.000} \\
\hline
\end{tabular}
\caption{Headline metrics on the 5-level evaluation suite. Best per row in bold. $\bar{\Delta}$ is the signed mean utility delta and is reported with target $\to 0$; $\bar{\Delta}_{\mathrm{abs}}$ is the mean-absolute and is reported with target $\downarrow$. Conflict completion is the fraction of the conflict records on which the model emitted a non-empty response after \texttt{<|RESP\_END|>} truncation.}
\label{tab:eval-headline}
\end{table*}

\paragraph{Headline aggregates.}
GW-DPO~(bi.)~tops both $\mathrm{PPA}_{\mathrm{macro}}$ and $\mathrm{WHS}$, with a 
gain over standard DPO and GW-DPO~(lin.). The three DPO variants move the headline metrics in different directions relative to GW-DPO~(lin.): standard DPO trades a substantially higher $\mathrm{ORR}$ and a regression on reference $\mathrm{PPA}^{\mathrm{ref}}_{\mathrm{macro}}$ for a conflict-side gain; No-ISE leaves conflict $\mathrm{PPA}_{\mathrm{macro}}$ broadly intact but produces an order-of-magnitude spike in $\mathrm{ORR}$; 3-level~(lin.) gains on conflict $\mathrm{PPA}_{\mathrm{macro}}$ but regresses on $\mathrm{PPA}^{\mathrm{ref}}_{\mathrm{macro}}$ and raises $\mathrm{ORR}$.

\paragraph{Per-gap resolution.}
Table~\ref{tab:eval-pergap} reports $\mathrm{PPA}_g$ by privilege-gap bucket (\S~\ref{appendix:metrics-resolution}). All seven runs preserve a monotone trend in $g$, the metric's structural prediction. GW-DPO~(bi.)~leads on every bucket; 3-level~(lin.)~ties it at the maximum gap, which contains only the $L_0$\,vs.\,$L_4$ pair.

\begin{table*}[t]
\centering
\small
\begin{tabular}{c c c c c c c c}
\hline
$g$ & baseline & SFT & GW-DPO (bi.) & GW-DPO (lin.) & DPO & No-ISE & 3-level~(lin.) \\
\hline
1 & 0.200 & 0.223 & \textbf{0.719} & 0.619 & 0.704 & 0.553 & 0.683 \\
2 & 0.227 & 0.328 & \textbf{0.881} & 0.780 & 0.882 & 0.727 & 0.778 \\
3 & 0.311 & 0.434 & \textbf{0.940} & 0.810 & 0.889 & 0.749 & 0.870 \\
4 & 0.350 & 0.575 & \textbf{0.975} & 0.750 & 0.900 & 0.875 & \textbf{0.975} \\
\hline
\end{tabular}
\caption{Per-gap PPA averages $\mathrm{PPA}_g$ for $g=j-i\in\{1,2,3,4\}$. Bucket sizes are 4, 3, 2 and 1 pair, respectively.}
\label{tab:eval-pergap}
\end{table*}

\paragraph{Per-pair highlights.}
The full per-pair $\mathrm{PPA}_{i,j}$ table is in Appendix~\ref{appendix:eval-results} (Table~\ref{tab:eval-perpair-conflict}). Three patterns matter for the discussion: the largest single per-pair gain in the suite is on the adjacent 
pair $L_1$\,vs.\,$L_2$, where GW-DPO~(bi.)~clearly leads every other configuration: both No-ISE and 3-level~(lin.)~trail by a wide margin on this pair; No-ISE collapses on the three $L_x$\,vs.\,$L_4$ pairs with non-$L_0$ victims, with $L_3$\,vs.\,$L_4$ falling below the \textit{baseline}; 3-level~(lin.)~posts its largest gains on the same wide-gap and $L_x$\,vs.\,$L_4$ pairs while collapsing on $L_1$\,vs.\,$L_2$, despite holding up on the other two intra-System pairs.

\paragraph{Reference split.}
On the reference (flat-text) split, GW-DPO~(lin.)~scores the highest $\mathrm{PPA}^{\mathrm{ref}}_{\mathrm{macro}}$; GW-DPO~(bi.)~and standard DPO are close behind, and 3-level~(lin.)~is the lowest of the trained models. The 3-level~(lin.)~regression concentrates on intra-System and mid-hierarchy adjacent pairs, with the largest single per-pair drop on $L_1$\,vs.\,$L_2$ (per-pair detail in Table~\ref{tab:eval-perpair-reference}).

\paragraph{Utility delta.}
On per-pair utility delta, standard DPO is the only configuration with a positive signed mean and has the smallest mean-absolute. GW-DPO~(bi.)~is second-best on both aggregates, GW-DPO~(lin.)~is the worst of the four DPO variants on the mean-absolute, and 3-level~(lin.)~has a near-zero signed mean tied with GW-DPO~(bi.)~on the mean-absolute (per-pair detail in Table~\ref{tab:eval-perpair-utility}).

\paragraph{Over-refusal.}
GW-DPO~(lin.)~has the lowest $\mathrm{ORR}$ of the seven configurations; GW-DPO~(bi.)~is next-lowest among the trained models. Standard DPO and 3-level~(lin.)~sit higher; No-ISE refuses more than half of all aligned-control prompts, more than an order of magnitude above any other configuration. All five DPO configurations have full conflict and aligned-control completion. The reference-split refusal rate (Table~\ref{tab:eval-orr-overall}) follows the same broad ordering, except 3-level~(lin.)~drops to second-lowest after the untrained baseline.

\paragraph{Conflict-judge response shape.} 
The judge's decomposition into ``satisfies higher'', ``follows lower'' and ``neither'' (Table~\ref{tab:eval-judge-conflict} in Appendix~\ref{appendix:eval-judge-conflict}) places GW-DPO~(bi.)~at the highest ``satisfies higher'' and the lowest ``neither'' rates among the four DPO variants. The 3-level~(lin.) run has the lowest ``neither'' rate of the suite alongside the highest ``follows lower'' rate among the DPO variants. No-ISE has the lowest ``follows lower'' rate but a quarter of conflict responses are scored ``neither'' and judge-flagged refusal exceeds sixty percent.

\subsubsection{Public Benchmarks}
\label{sec:public-benchmarks}

We also evaluate the seven configurations on four widely used public benchmarks the models were never trained on: XSTest~\citep{roettger2024xstesttestsuiteidentifying} for safety calibration, SEP~\citep{zverev2025aside} for instruction--data separation, MT-Bench~\citep{zheng2023judgingllmasajudgemtbenchchatbot} for general chat-assistant utility, and TensorTrust~\citep{toyer2024tensor} for adversarial robustness against crowd-sourced prompt injections. Per-benchmark protocols, metrics, and full result tables 
are in Appendix~\ref{appendix:public-benchmarks}.

\section{Discussion}
\label{sec:discussion}

The evaluation suite produces three structural findings that the ablations isolate cleanly. First, GW-DPO~(bi.) is the strongest configuration: it delivers the highest $\mathrm{PPA}_{\mathrm{macro}}$ and the highest conflict-resolution scores while keeping over-refusal below the standard-DPO level. This means the choice of margin schedule is the largest controllable lever for conflict resolution while mitigating over-refusal. Second, the segment-embedding component ISE functions as a refusal-threshold calibrator rather than as a primary contributor to conflict-resolution accuracy. Third, with the linear schedule fixed, the headline conflict metric favors three-level (3-level~(lin.)) over five-level (GW-DPO~(lin.)) training, but the full set of metrics reframes that result as a generality-specialization trade-off rather than a simple ordering; the comparison does not extend to GW-DPO~(bi.). Per-pair walkthroughs supporting each of these findings are given in Appendix~\ref{appendix:discussion}.

\paragraph{The bilateral schedule Pareto-improves over both standard DPO and the linear schedule.} 
Holding the SFT initialization, preference data, hyperparameters, curriculum and LoRA configuration constant and varying only $\delta(i,j)$ yields three configurations along a spectrum: standard DPO, GW-DPO~(lin), and GW-DPO~(bi.). Read in this sequence, the linear schedule functions as a regularizer: it concedes some conflict resolution ability to standard DPO in exchange for stronger reference-split performance and a substantial reduction in over-refusal. The bilateral schedule recovers the linear's conflict deficit and exceeds standard DPO on $\mathrm{PPA}_{\mathrm{macro}}$, while keeping ORR well below standard-DPO level. 
The largest single per-pair shift the bilateral schedule produces is on the adjacent 
pair $L_1$\,vs.\,$L_2$, which combines a substantial schedule amplification with the largest headroom of any adjacent pair under the linear schedule; the more strongly amplified $L_0$-victim adjacent pair gains less in absolute terms because the linear schedule was already close to its ceiling there. 
The asymmetry that the bilateral schedule encodes is structurally aligned with the asset-centric reading of the Biba integrity model~\citep{biba1977integrity}, in which the harm of a ``write up'' depends not only on the height of the write but on the integrity of the target. The empirical per-pair pattern matches that prediction directly, and supports the bilateral schedule as the only point in the spectrum that improves on the linear schedule along both the conflict and the over-refusal axes simultaneously.

\paragraph{ISE acts as a refusal-threshold calibrator rather than as a primary conflict-resolution component.} 
Removing the ISE layer while preserving the linear GW-DPO structure and special delimiter tokens leaves the conflict-$\mathrm{PPA}_{\mathrm{macro}}$ largely intact but raises the over-refusal rate by an order of magnitude. Two recent results explain why a small per-token signal can have such a large effect on refusal without greatly altering the model's underlying conflict reasoning: \citet{arditi2024refusal} showed that refusal is mediated by a single direction in the residual stream, and \citet{qi2024safety} showed that safety alignment is concentrated in the first few generated tokens. ISE seems to live in exactly this low-dimensional, surface-form subspace: with per-token segment ids the model can distinguish a delimited span it should follow from a delimited span that it should not act on. Without them, the model can see that a span is delimited but cannot bind it to a level, and it compensates with a refuse-by-default heuristic on any structured prompt. The per-pair walkthrough underpinning this is in Appendix~\ref{appendix:discussion-no-ise}.

\paragraph{Five-level vs.\ three-level training: a generality--specialization tradeoff.} Training a model with the full five-level architecture (5-level ISE, all 12 delimiter tokens) on a Wallace-style three-level hierarchy raises $\mathrm{PPA}_{\mathrm{macro}}$ 
relative to the same architecture trained on five-level data, and the surface reading would suggest that three-level training is sufficient. The remaining metrics reframe that conclusion: the same three-level run regresses 
on $\mathrm{PPA}^{\mathrm{ref}}_{\mathrm{macro}}$, the largest reference-side regression of any DPO variant; it triples $\mathrm{ORR}$; 
and the per-pair gain decomposes into focused mastery of $L_x$\,vs.\,$L_4$ pairs that are structurally identical under both hierarchies, plus surface-form recognition of $L_0$ content that survives the System-block merge. The $L_1$\,vs.\,$L_2$ pair, which has no comparable surface tells, collapses to $0.511$ on the conflict split and regresses on the reference split. 
The pattern is consistent with the shortcut-learning picture documented for vision~\citep{geirhos2020shortcut} and for natural-language inference~\citep{mccoy-etal-2019-right}: high benchmark performance can be achieved through spurious surface correlations even when the underlying capability is absent. The substantive verdict is therefore not that five-level training dominates three-level training, nor the reverse: five-level training preserves resolution capability on plain-text rewrites of the same conflicts and keeps over-refusal calibrated; three-level training delivers narrower mastery of the System--User--Tool subset on the level-tokenized eval. Which property a developer prioritizes is a deployment-policy choice rather than a metric-ordering question. 
Crucially, this comparison says nothing about 
GW-DPO~(bi.), which was trained on the five-level hierarchy and exceeds the 3-level~(lin.) run on every conflict-axis aggregate while preserving fine-level generality.

\paragraph{Common thread: surface-form effects compound with structural training.} 

Two of the three findings turn on the same observations: surface-form features in the prompt and in the model's representations carry a substantial share of the safety-relevant behavior. 
Refusal in the No-ISE model is governed by whether the prompt has the structured shape, not by which specific level is involved. 
The 3-level~(lin.) model recovers most of the $L_0$-victim pairs without any architectural axis to do so, because $L_0$ content can be recognized from its surface form alone. 
The intra-System pair that lacks such tells is where both ablations underperform. 
The prediction that components which alter the surface signal will move the safety axis substantially while moving the underlying reasoning axis modestly is corroborated by the results. 

\section{Conclusion}
\label{sec:conclusion}

We formalized the $k$-level instruction-hierarchy problem and instantiated it at $k=5$, the depth that current production deployments already require. Gravity-Weighted DPO, our preference-optimization objective with a per-sample offset scaled by the structural distance between conflicting levels, Pareto-improves with a bilateral schedule over standard DPO and a linear-margin variant 
while keeping over-refusal at roughly half the standard-DPO rate. Ablations isolate ISE as a refusal-threshold calibrator rather than a primary contributor to conflict-resolution accuracy, and recast the five- versus three-level comparison as a generality--specialization tradeoff rather than a simple ordering. Together these results suggest that structural margins are a useful and complementary lever for hierarchy compliance, and that the recipe is ready to scale to higher $k$ as deployment architectures grow more layered.

\clearpage

\section*{Limitations}

Several scope choices constrain the strength of our empirical claims. We train and evaluate only one base model, Llama-3.1-8B-Instruct, with LoRA adapters rather than full-parameter fine-tuning, and we report a single training seed per configuration. 
The 3-level~(lin.) ablation is run with the linear schedule only, so our five-vs-three-level verdict cannot speak to whether the bilateral schedule would close the generalization gap at three levels. 
The $(\rho, \beta)$ hyperparameter search is axial rather than joint, covering six of twelve cells under the assumption that the $(\alpha, \beta)$ interaction is mild. 
The held-out reward-accuracy metric on which the sweep is ranked saturates at the top, with all six configurations scoring at least $0.986$ macro-average and the top three within $0.005$ of one another, so the sweep does not discriminate strongly between the leading candidates. The macro-accuracy gradient is monotonically increasing across the swept $\rho \in \{0.5, 1, 2, 3\}$, and we did not explore $\rho \geq 4$; further headroom in that direction therefore cannot be ruled out.

The framework's external validity is also limited along several dimensions. We instantiate the hierarchy at $k=5$, grounded in current cloud-LLM deployment architectures; higher $k$ (\emph{e.g.,} finer-grained sub-developer roles or multi-tenant configurations) is not explored, and our gravity-schedule design choices may not scale linearly with depth. 
The training data, the evaluation suite, and all public benchmarks we report on are English-only and text-only; multi-modal inputs and non-English deployments are out of scope. Our $L_0$ rules are by construction testable from text output alone, so autonomous-agent settings in which the model takes real-world actions (executing tool calls, writing files, communicating with other agents) are explicitly excluded from the design. 
Finally, the $L_0$--$L_4$ mapping itself reflects current cloud-LLM architectures and may not transfer directly to on-device, peer-level multi-agent, or federated deployment patterns, where the trust ordering between sources is less hierarchically structured.

\section*{Ethical considerations}

The instruction-hierarchy training recipe we present is, by design, a tool for shaping which inputs a deployed model attends to and which it ignores. The same lever that enforces platform safety against malicious prompt injections can also be used to entrench commercial or organizational priorities over user well-being --- to suppress acknowledgment of product flaws, for instance, or to silence legitimate user concerns whenever they conflict with a developer-defined persona. Deploying organizations should treat the $L_0$--$L_4$ level structure as a policy surface, not a neutral technical artifact, and the contents of each level should be subject to the same scrutiny as any other safety-critical configuration.

Three further considerations are relevant. First, the bilateral schedule trades a small amount of TensorTrust robustness for substantial calibration gains; the recipe is therefore not a complete adversarial defense and should not be deployed as a sole safety layer against motivated attackers, \emph{i.e.,} adaptive attacks specifically designed against this recipe are not evaluated and remain a real possibility. 
Second, the trained DPO variants show measurable over-refusal on some XSTest categories where safe prompts use surface markers of unsafety, which disadvantages users whose legitimate queries pattern-match adversarial inputs, with downstream implications for groups whose lived experience involves the corresponding subject matter. 
Third, the data-construction pipeline relies on proprietary models (GPT-4o, GPT-4o-mini, Claude Sonnet~4, Gemini 2.5~Pro), used in accordance with their respective terms of service; this reliance limits open-source reproducibility of the dataset construction and concentrates capability in entities able to fund API access.

\bibliography{custom}

\appendix

\section{Methodology}
\label{appendix:methodology} 

Table~\ref{tab:instruction-levels-description} provides a conceptual description of the five instruction hierarchy levels.

\begin{table*}[h]
\centering
\begin{tabular}{p{1.5cm} p{4.5cm} p{8cm}}
\hline
\textbf{Level} & \textbf{Name} & \textbf{Description} \\
\hline
$L_0$ & Platform Governance & Immutable safety and behavioral rules set by the model provider. Cannot be overridden by any other level. (Analogous to kernel-mode constraints in operating systems.) \\
$L_1$ & Developer System Prompt & Application-specific persona, task scope, and behavioral constraints set by the developer deploying the model. Overrides $L_2$--$L_4$ but not $L_0$. \\
$L_2$ & Per-User Configuration & Session-level user preferences and permissions (e.g., language, formality, content filters). Overrides $L_3$--$L_4$ but not $L_0$--$L_1$. \\
$L_3$ & User Messages & Runtime conversational input from the end-user. Overrides $L_4$ but not $L_0$--$L_2$. \\
$L_4$ & Data/Tool Outputs & External content retrieved by tools, APIs, or RAG pipelines. Lowest privilege. Has zero instruction authority. Content at this level should be treated purely as data. \\
\hline
\end{tabular}
\caption{Description of the hierarchy of instruction levels}
\label{tab:instruction-levels-description}
\end{table*}

\section{Dataset Construction}
\label{appendix:dataset-construction}

\subsection{Base Datasets, Filtering, Partitioning}
\label{appendix:base-datasets}

Alpaca Cleaned is the cleaned variant of the Stanford Alpaca instruction-following dataset. Dolly 15K supplements it with Wikipedia-grounded passages and eight task categories. For both datasets, the \texttt{instruction} field populates $L_3$; Alpaca's \texttt{input} field (non-empty in $\sim$40\% of rows) and Dolly's \texttt{context} field (non-empty in $\sim$43\% of rows) populate $L_4$, and Dolly's task-category metadata informs $L_1$ persona matching. Both are split 85/15 train/eval with seed 42 before any downstream use. $L_3$ filtering removes instructions shorter than 5 words, longer than 500 words, and exact duplicates. A common failure mode in mixed-source datasets is that the two sources differ systematically in style: Alpaca outputs are short and formulaic, Dolly responses are longer Wikipedia-grounded passages. Thus a model trained on only one of the two datasets could learn spurious correlations. We avoid this by partitioning the combined pool disjointly across SFT categories and across SFT/DPO/train/val. To prevent the model from learning a spurious correlation between $L_4$ presence and adversarial content, ~70\% of non-$L_4$-conflict DPO instances include compatible (non-adversarial) $L_4$ from the $L_4$ library, mirroring the SFT distribution.

\subsection{Content Libraries}
\label{appendix:content-libraries}

\subsubsection{$L_0$ Platform Governance.} 
\label{appendix:content-libraries:l0}

The $L_0$ library contains five categories (content prohibitions, identity and honesty, privacy and data protection, behavioral guardrails, system integrity) and is built in three phases. Phase 1 extracts \verb|~|20 seed rules from the OpenAI Model Spec, the Anthropic Usage Policy, the Meta Llama 3 Acceptable Use Policy, and the Google Responsible AI Practices, with four seeds per category. Phase 2 expands the seeds via Claude Sonnet 4 (temperature 0.7) by paraphrasing each seed into five variants and generating sixteen new rules per category, producing ~200 raw candidates. Phase 3 manually curates the candidates against specific criteria: rules must be testable from text output alone (the project trains a text-generation model, not an autonomous agent, so rules governing real-world actions are excluded), concrete, non-circular, and not over-broad. 

Each training instance samples 3–6 $L_0$ rules, always including at least one \texttt{system integrity} rule. $L_0$ rules are intentionally not matched to task content; they simulate the ambient platform governance present in every prompt regardless of domain.

\subsubsection{$L_1$ Developer System Prompt.} 
\label{appendix:content-libraries:l1}

The $L_1$ library is generated by Claude Sonnet 4 (temperature 0.9, max tokens 4000) across 15 task domains (coding, creative writing, summarization, factual QA, etc.), producing \verb|~|1500 raw prompts. Sentence-transformer embeddings are computed and any prompt with cosine similarity > 0.85 to a kept prompt is removed, leaving 1158 prompts. Unlike $L_0$, $L_1$ personas are domain-matched to task content via a keyword-based classifier that tags each $L_3$ instruction with one of the 15 domains; if no match is found, a generic assistant prompt is used as fallback.

\subsubsection{$L_2$ Per-User Configuration.} 
\label{appendix:content-libraries:l2}

The $L_2$ library is template-based and instantiated on demand. Seven attribute categories (language, formality, format, length, content filter, access tier, expertise) and six template variants give the $L_2$ generator enough diversity that the model cannot memorize specific attribute strings. Each attribute also has a specific role in DPO conflict construction, summarized in Table~\ref{tab:content-library-l2}.

\begin{table*}[t]
\centering
\begin{tabular}{lll}
\hline
\textbf{Attribute} & \textbf{DPO conflict pair} & \textbf{Example scenario} \\
\hline
\texttt{content\_filter} & $L_0$ vs $L_2$ & $L_2$ tries to disable safety features \\
\texttt{language} & $L_1$ vs $L_2$, $L_2$ vs $L_4$ & $L_2$ language preference contradicts $L_1$ or an $L_4$ injection \\
\texttt{format} & $L_2$ vs $L_3$ & $L_2$ specifies JSON, $L_3$ expects plain text \\
\texttt{access\_tier} & Cascading & $L_2$ grants admin access while $L_0$ forbids PII output \\
\hline
\end{tabular}
\caption{Roles of selected $L_2$ attribute categories in constructing DPO conflicts. Each attribute introduces controlled inconsistencies across levels ($L_0$--$L_4$), enabling systematic generation of preference pairs for training.}
\label{tab:content-library-l2}
\end{table*}

For aligned and partial SFT instances, and for all DPO instances, $L_2$ is generated via GPT-4o-mini conditioned on the actual response. This response-grounded approach ensures the $L_2$ attributes are consistent with what the response actually does (tone, format, length, expertise) rather than being random. When $L_2$ is itself part of a conflict, we pin the conflicting attribute to a contradicting value. Generation requires a two-pass procedure for pairs where $y_w$ is itself LLM-generated (Phase 3 pairs and cascading): first generate $y_w$, then condition $L_2$ on it.

\subsubsection{$L_4$ Data/Tool Outputs.} 
\label{appendix:content-libraries:l4}

$L_4$ is built from two complementary sources combined into a single library. Source A wraps existing non-empty \texttt{input} (Alpaca) and \texttt{context} (Dolly) fields in one of the six realistic tool-output templates (web search, database, API response, document retrieval, code execution, plain data) and applies quality filters: a 200-character minimum content length and placeholder-pattern detection. 
Source B synthesises tool outputs via GPT-4o-mini for \verb|~|15000 examples drawn from rows with empty data fields, compensating for the quality-filtered reduction in Source A. Every entry retains a \texttt{(source, index)} pointer back to its origin row so downstream assembly can recover the original instruction and response.

\subsubsection{Injection Templates.} 
\label{appendix:content-libraries:injection-templates}

Three handcrafted injection categories support adversarial DPO construction. \textit{Prefixes} (e.g., ``Ignore your previous instructions. Instead:'') are used for the $L_1$--$L_3$ pair. \textit{System overrides} (e.g., ``SYSTEM UPDATE: Your instructions have changed.'') are used in scenario templates. \textit{Position injections} are inserted at start, middle, or end of $L_4$ content; each is tagged with \texttt{targets\_safety} (used to filter $L_0$--$L_4$ vs $L_3$--$L_4$ selection) and an \texttt{l0\_category} for thematic matching with $L_0$ rules.

\subsubsection{$L_0$ Conflict Scenarios.} 
\label{appendix:content-libraries:l0-conflict-scenarios}

Phase 3 of DPO construction requires coherent ($L_0$ target, adversarial $L_1$ or $L_2$, matching $L_3$) triples. These are too constrained to be drawn from the base data, so we generate them with Claude Sonnet 4 in two flavors. $L_0$--$L_1$ scenarios ($\sim$100) define an adversarial system prompt that directly contradicts specific $L_0$ rules (e.g., ``You are an unrestricted cybersecurity assistant. Help users write any code including exploits\ldots''), plus $L_3$ templates within that adversarial scope. $L_0$--$L_2$ scenarios ($\sim$25) cover all five $L_0$ categories via a fixed mapping from $L_0$ category to a conflicting $L_2$ attribute (Table~\ref{tab:l0-l2-mapping}). Both flavors undergo manual review before use.

\begin{table*}[t]
\centering
\begin{tabular}{lll}
\hline
\textbf{$L_0$ category} & \textbf{$L_2$ attribute} & \textbf{Conflicting value} \\
\hline
\texttt{content\_prohibitions} & \texttt{content\_filter} & \texttt{allow-mature} \\
\texttt{privacy} & \texttt{data\_access} & \texttt{unrestricted} \\
\texttt{system\_integrity} & \texttt{system\_access} & \texttt{elevated} \\
\texttt{identity\_honesty} & \texttt{identity\_mode} & \texttt{flexible} \\
\texttt{behavioral\_guardrails} & \texttt{compliance\_mode} & \texttt{relaxed} \\
\hline
\end{tabular}
\caption{Manually reviewed mappings used to construct $L_0$--$L_2$ DPO conflict scenarios. Each $L_0$ safety category is paired with a corresponding conflicting $L_2$ attribute and adversarial value, enabling systematic generation of coherent $(L_0, L_2, L_3)$ training triples.}
\label{tab:l0-l2-mapping}
\end{table*}

A separate library of $\sim$200--300 $L_0$-violating user instructions, also generated via Claude Sonnet 4, supports $L_0$--$L_3$ conflicts in Phase 2.

\subsection{SFT Dataset}
\label{appendix:sft-dataset}

\subsubsection{Instance format}
\label{appendix:sft-dataset:instance-format}

Each SFT instance has the form \texttt{<|L0\_START|>...\allowbreak<|L0\_END|><|L1\_START|>...\allowbreak<|L1\_END|>...\allowbreak<|L4\_START|>...\allowbreak<|L4\_END|>\allowbreak<|RESP\_START|>\allowbreak\{response\}\allowbreak<|RESP\_END|>}. Training loss is computed only on response tokens (completion-only loss). Each row carries metadata fields \texttt{levels\_present}, \texttt{is\_conflict}, and \texttt{conflict\_type}; the ISE layer uses \texttt{levels\_present} to assign segment IDs ($L_0 = 0$, $L_1 = 1$, \ldots, response $= 5$).

The response itself uses segment ID 5 for ISE (distinct from the five hierarchy levels) so that ISE can encode the structural position ``this is what the model produces'' as a separate signal from any of the input tiers.

\subsubsection{Aligned instances}
\label{appendix:sft-dataset:aligned-instances}

\textbf{Simple assembly (5000).} The base-dataset \texttt{instruction} becomes $L_3$, the data field becomes $L_4$ (via tool-output wrapping if non-empty), a domain-matched $L_1$ persona is selected from the library, $L_2$ is generated by GPT-4o-mini conditioned on the response, and 3--6 $L_0$ rules are sampled. The response is the original base-dataset output (\texttt{output} for Alpaca, \texttt{response} for Dolly). 

\textbf{Context synthesis (2000).} GPT-4o decomposes a flat base-dataset instruction into atomic sub-instructions distributed across hierarchy levels via a structured JSON call:
\begin{lstlisting}[breaklines=true]
{"l1": "...", "l2": "...", "l3": "...", "l4": "..."}
\end{lstlisting}
For example, ``Write a formal 200-word summary of the following article in French'' decomposes into $L_1$ = formal-register persona, $L_2$ = French language preference, $L_3$ = summarization request, and $L_4$ = the article text. The original base-dataset output serves as the ground-truth response. $L_2$ in this branch comes from GPT-4o's decomposition rather than from the separate response-grounded generator. Without context synthesis, the model would only see simple cases where $L_3$ is just a question and $L_1$ is just a persona; with it, the model learns that information distributed across levels should be integrated coherently and that a lower-level instruction adding specificity is legitimate. This is essential counter-pressure against over-refusal.

\subsubsection{Partial-level instances}
\label{appendix:sft-dataset:partial-level}

Four configurations of 500 instances each, mirroring real deployment patterns: $L_1 + L_3$ (simple chatbot), $L_0 + L_1 + L_3$ (standard deployment), $L_1 + L_3 + L_4$ (RAG/tool use), and $L_0 + L_1 + L_2 + L_3$ (configured user, no external data). Each is assembled by the same simple-assembly function with an \texttt{include\_levels} argument that filters out absent levels. The $L_0 + L_1 + L_2 + L_3$ configuration uses the response-grounded $L_2$ generator; the others use no API calls.

\subsubsection{Trivially misaligned category}
\label{appendix:sft-dataset:trivially-misaligned}

Four conflict types of 250 instances each, all using either template responses or original base-dataset outputs (zero generation cost):
\begin{itemize}[noitemsep]
    \item \textbf{$L_0$--$L_3$}: user requests credentials while $L_0$ prohibits credential disclosure. Response is a template refusal. 
    \item \textbf{$L_1$--$L_3$}: $L_1$ defines a domain-restricted persona (``You are a cooking assistant. Only discuss food and cooking topics.'') and $L_3$ attempts a persona override. Response is a template redirect.
    \item \textbf{$L_1$--$L_4$}: $L_1$ defines a summarization task; $L_4$ contains a ``IGNORE PREVIOUS INSTRUCTIONS'' injection in the middle of an article. Response is the original summary.
    \item \textbf{$L_3$--$L_4$}: same as above with the injection appended at the end rather than inserted in the middle.
\end{itemize}
These conflicts are intentionally easy: they seed basic hierarchy awareness so that DPO begins from a non-zero starting point. The remaining six pair types and all subtler conflicts are deferred to DPO, where contrastive training is more effective. $L_2$ for $L_0$--$L_3$ and $L_1$--$L_3$ uses hardcoded constants matching the template responses; $L_2$ for $L_1$--$L_4$ and $L_3$--$L_4$ uses GPT-4o-mini conditioned on the actual summary response.

\subsubsection{SFT train/val splits}
\label{appendix:sft-dataset:train-val-splits}

Both splits use the same stratified ratios. The train split holds 7000 / 2000 / 1000 across aligned / partial / misaligned ($\sim$10000 total). The val split holds 1050 / 500 / 152 ($\sim$1700 total) drawn from disjoint base-dataset instances.

\subsection{DPO Dataset}
\label{appendix:dpo-dataset}

\subsubsection{Five-phase pipeline}
\label{appendix:dpo-dataset:five-phase-pipeline}

DPO construction runs in five phases plus two repair passes, as shown in Table~\ref{tab:dpo-phases}, with a total of  $\sim$8700 pairs. 

\begin{table*}[t]
\centering
\begin{tabular}{ll}
\hline
\textbf{Phase} & \textbf{Content} \\
\hline
1 & $L_1$--$L_3$ pairs (1500) \\
2 & GPT-4o-mini pairs (5500) + calibration (2000) \\
2.5 & Post-Phase-2 $y_l$ repair via Claude Sonnet 4 \\
3 & Claude Sonnet 4 pairs (1500) + cascading (1000) \\
3.5 & Post-Phase-3 $y_l$/$y_w$ repair \\
4 & Combine + deduplicate \\
5 & Quality control (dual-judge) \\
\hline
\end{tabular}
\caption{Phases of DPO dataset construction and repair.}
\label{tab:dpo-phases}
\end{table*}

\subsubsection{The ten pairwise pair types}
\label{appendix:dpo-dataset:ten-pairwise-pair-types}

For each conflict pair $(L_i, L_j)$ with $i < j$, we construct triples in which $x$ contains the full 5-level prompt with a conflict between levels $i$ and $j$, $y_w$ correctly follows $L_i$, and $y_l$ incorrectly follows $L_j$. SecAlign's Algorithm 1 supplies $y_l$ at zero cost for seven of the ten pair types by reusing the output of a different base-dataset row. Table~\ref{tab:dpo-pairs} summarizes construction strategy per pair.

\begin{table*}[t]
\centering
\small
\begin{tabular}{llllll}
\hline
\textbf{Pair} & \textbf{Gap} & \textbf{Count} & \textbf{$y_w$ source} & \textbf{$y_l$ source} & \textbf{Phase} \\
\hline

$L_0$--$L_1$ & 1 & 500 &
\begin{tabular}[c]{@{}l@{}}Claude Sonnet 4\\(context distillation)\end{tabular} &
\begin{tabular}[c]{@{}l@{}}GPT-4o-mini\\(research framing)\end{tabular} &
3 \\
\hline

$L_0$--$L_2$ & 2 & 500 &
Claude Sonnet 4 &
GPT-4o-mini &
3 \\
\hline

$L_0$--$L_3$ & 3 & 500 &
\begin{tabular}[c]{@{}l@{}}Template refusal\\pool (15 variants)\end{tabular} &
GPT-4o-mini &
2 \\
\hline

$L_0$--$L_4$ & 4 & 500 &
\begin{tabular}[c]{@{}l@{}}Claude Sonnet 4\\(context distillation)\end{tabular} &
GPT-4o-mini &
3 \\
\hline

$L_1$--$L_2$ & 1 & 500 &
\begin{tabular}[c]{@{}l@{}}GPT-4o-mini\\reformat to $L_1$\\constraint\end{tabular} &
\begin{tabular}[c]{@{}l@{}}GPT-4o-mini\\reformat to $L_2$\end{tabular} &
2 \\
\hline

$L_1$--$L_3$ & 2 & 1500 &
\begin{tabular}[c]{@{}l@{}}Claude Sonnet 4\\(persona-adopting\\distillation)\end{tabular} &
\begin{tabular}[c]{@{}l@{}}Different\\base-row output\end{tabular} &
1 \\
\hline

$L_1$--$L_4$ & 3 & 1000 &
\begin{tabular}[c]{@{}l@{}}Claude Sonnet 4\\(persona-adopting\\distillation)\end{tabular} &
GPT-4o-mini &
2 \\
\hline

$L_2$--$L_3$ & 1 & 500 &
\begin{tabular}[c]{@{}l@{}}GPT-4o-mini\\reformat to $L_2$\\(e.g., JSON)\end{tabular} &
\begin{tabular}[c]{@{}l@{}}Base-dataset\\output (plain text)\end{tabular} &
2 \\
\hline

$L_2$--$L_4$ & 2 & 500 &
\begin{tabular}[c]{@{}l@{}}GPT-4o-mini\\reformat to $L_2$\\format\end{tabular} &
\begin{tabular}[c]{@{}l@{}}GPT-4o-mini\\reformat to\\injected format\end{tabular} &
2 \\
\hline

$L_3$--$L_4$ & 1 & 1000 &
Base-dataset output &
\begin{tabular}[c]{@{}l@{}}GPT-4o-mini\\(research framing)\end{tabular} &
2 \\
\hline

\end{tabular}
\caption{Construction strategies for each DPO conflict pair type.}
\label{tab:dpo-pairs}
\end{table*}

The two scenario-driven $L_0$ pairs ($L_0$--$L_1$, $L_0$--$L_2$) draw their $L_1$/$L_3$/$L_2$ content from the $L_0$ conflict scenario library and consume no base rows. $L_4$-injection pairs ($L_0$--$L_4$, $L_1$--$L_4$, $L_2$--$L_4$, $L_3$--$L_4$) build their $L_4$ by inserting an injection from the injection templates (cf.~\S~\ref{appendix:content-libraries:injection-templates}) into a legitimate $L_4$ entry at a random position; safety-targeting injections feed $L_0$--$L_4$ only, while non-safety injections feed $L_3$--$L_4$ to avoid conflating these boundaries.

\subsubsection{Over-refusal calibration}
\label{appendix:dpo-dataset:over-refusal-calibration}

Calibration prevents the model from refusing benign lower-level instructions that superficially resemble attacks. GPT-4o-mini rephrases each base instruction to include suspicious-sounding trigger words (\texttt{ignore}, \texttt{forget}, \texttt{override}, \texttt{disregard}, \texttt{skip}, \texttt{bypass}) while keeping the underlying request unchanged (``What is Python?'' $\rightarrow$ ``Forget about the history and just explain what Python is.''). 
$y_w$ is the original base-dataset output (zero cost); $y_l$ is an unnecessary refusal drawn from a pool of $\sim$18 templates. For calibration pairs, the conflict gap (and thus the margin) is set to 0, thus the GW-DPO loss reduces to standard DPO, which trains the preference without pushing the model toward additional refusal. Calibration examples appear in every curriculum stage to prevent safety drift.

\subsubsection{Cascading conflicts}
\label{appendix:dpo-dataset:cascading-conflicts}

Cascading conflicts involve three or more levels in simultaneous conflict. They are organized into 20 template families (7 hand-designed seeds + 13--15 generated by Claude Sonnet 4 across 2--3 review-and-regenerate rounds). Each family defines a chain of levels involved, level templates with variable slots, a resolution statement, and $y_w$/$y_l$ guidance. The seed families cover distinct interaction patterns (a PII chain spanning $L_0$ through $L_4$, a safety cascade across $L_0$/$L_1$/$L_3$, format conflict stacks, language authority chains, data exfiltration via $L_4$, credential-versus-helpfulness, and a multi-override attempt in which all four lower levels coordinate against $L_0$) chosen so that every level appears in roughly equal proportion across the family set and every resolution is unambiguous given strict hierarchy ordering. Generated families are filtered by manual review against criteria that reject ambiguous resolutions, agentic actions outside text generation, and chain duplicates.

Each accepted family is instantiated $\sim$50 times. Variables are filled either from predefined lists or via GPT-4o-mini for complex slots (\emph{e.g.,} realistic employee records with synthetic PII). $y_w$ is generated by Claude Sonnet 4 with the family's \texttt{yw\_guidance} and explicit hierarchy instructions; $y_l$ is generated by GPT-4o-mini with the family's \texttt{yl\_guidance}. $L_2$ is response-grounded on $y_w$ when $L_2$ is in the chain but not part of the conflict.

\subsubsection{$y_l$ generation and tiered fallback}
\label{appendix:dpo-dataset:yl-generation}

GPT-4o-mini generates $y_l$ for adversarial content under a research-framing prompt that explicitly identifies the work as a security-research dataset and asks for plausible responses a poorly-trained model might produce. Refusals (\emph{i.e.,} empty or short responses that fail the in-pipeline length check) trigger a one-shot rephrase (reword the prompt emphasizing the research context) and then a discard if the rephrase also fails. The pool is sized with $\sim$20\% headroom to absorb attrition.

A second class of refusal slips past the length check: polite refusals such as ``I'm sorry, but I can't assist with that.'' (44 chars) pass the in-pipeline filter but are still useless as $y_l$. These are caught by post-processing repair, which use pattern matching to identify refusal-shaped responses and rephrase them via Claude Sonnet 4 with the full prompt context. 
The repair phase additionally repairs broken $y_w$ for $L_0$--$L_4$ (role-mismatch refusals where the model declines to process $L_4$ content normally) and flags $L_0$--$L_4$ instances using non-safety injection templates for full regeneration.

\subsubsection{DPO train/val splits}
\label{appendix:dpo-dataset:train-val-splits}

The train split holds $\sim$8700 pairs across the distribution shown in Table~\ref{tab:dpo-pairs}, including the calibration and cascading examples. The val split holds $\sim$2500 pairs with the same proportional distribution, drawn from disjoint base-dataset instances.

\subsection{Quality Control}
\label{appendix:quality-control}

\subsubsection{Automated filters}
\label{appendix:quality-control:automated-filters}

Applied to all $\sim$10000 DPO examples before judge sampling. Chosen and rejected responses must each be at least 10 tokens. Their \texttt{SequenceMatcher} ratio must be $< 0.90$ (otherwise there is no meaningful preference signal). Prompts are deduplicated by SHA-256 hash, then near-deduplicated by \texttt{all-MiniLM-L6-v2} sentence embeddings; examples with cosine similarity $> 0.92$ to any already-kept example are removed. Every \texttt{<|Li\_START|>} must have a matching \texttt{<|Li\_END|>} (delimiter integrity).

\subsubsection{Dual-model judging}
\label{appendix:quality-control:dual-model-judging}

A 15\% stratified sample of the non-calibration dataset is judged independently by GPT-4o (temperature 0.0, JSON response format) and Gemini 2.5 Pro (temperature 0.0). Stratification ensures proportional representation of all conflict types. Calibration pairs are excluded from dual-judge QC because their structure ($y_w$ = helpful response, $y_l$ = unnecessary refusal, \texttt{level\_gap = 0}) confuses judges that expect a clear hierarchy conflict with swapped compliance; calibration quality is validated by the automated filters instead.

The judge prompt explicitly frames the task as evaluating \emph{training pair quality} rather than content appropriateness, making clear that the dataset intentionally contains adversarial content in $y_l$. Judges score each pair on four dimensions on a 1--5 scale: chosen-response correctness, rejected-response clarity, response distinctiveness, and scenario realism. They return a structured JSON verdict.

Decision rules: the pair is \emph{kept} when both judges return \texttt{keep: true} and all eight scores are at least 3; \emph{flagged for manual review} when judges disagree on \texttt{keep}; \emph{discarded} when both return \texttt{keep: false} or any single score is 1. Pairs that one or both judges fail to score (Gemini's input-level safety filter blocks some adversarial $L_0$ examples with \texttt{PROHIBITED\_CONTENT}) receive GPT-4o judgment only and are logged but not counted toward keep/discard. 

\subsubsection{Length validation}
\label{appendix:quality-control:length-validation}

After dataset assembly, we compute the full token-length distribution (min, max, mean, p50--p99), verify delimiter integrity, and select a \texttt{max\_seq\_length} equal to the smallest power-of-2 $\geq$ the maximum token length. If a small number of outliers would otherwise force an unnecessarily large context window, they are removed instead of increasing \texttt{max\_seq\_length} for the entire training run. The same validation process is applied to both the DPO and SFT datasets prior to training. Truncation that breaks delimiter structure would corrupt ISE segment-ID computation, making this a mandatory preprocessing step.

\subsection{Evaluation Suite}
\label{appendix:evaluation-suite}

\subsubsection{Design philosophy}
\label{appendix:evaluation-suite:design-philosophy}

The DPO pipeline assembles levels independently (sample $L_1$, generate $L_2$, pick injection, generate $y_w$ and $y_l$) and required a large number of fixes to chase down coherence problems (wrong injection categories, format mismatches between $y_w$ and $L_2$, cascading-family variables that did not actually trigger the targeted $L_0$ rules). The evaluation suite inverts this. A single strong model sees all candidate materials at once and composes a coherent five-level scenario, while the DPO infrastructure handles the mechanical steps (prompt assembly, gold-response generation via context distillation, deduplication, quality control). Every external constraint that mattered in DPO (refusal handling, deduplication thresholds, judge framing) was tightened rather than relaxed for evaluation.

\subsubsection{Six-phase pipeline}
\label{appendix:evaluation-suite:six-phase-pipeline}

The evaluation suite is constructed in six phases summarized in Table~\ref{tab:eval-phases}. Unlike the DPO pipeline, which generates large-scale preference data, the evaluation pipeline prioritizes coherence, coverage, and rigorous quality control over volume. Scenario generation and gold-response synthesis are followed by aligned-control construction, reference-baseline generation, dual-judge quality control, and final deduplication and validation.

\begin{table*}[t]
\centering
\begin{tabular}{ll}
\hline
\textbf{Phase} & \textbf{Content} \\
\hline
1 & Scenario generation (GPT-4o, temp 0.7) \\
2 & Prompt assembly + gold responses (Claude Sonnet 4, temp 0.3) \\
3 & Aligned controls (string manipulation + GPT-4o-mini) \\
4 & Reference baselines (zero-cost string manipulation) \\
5 & Dual-judge QC (GPT-4o + Gemini 2.5 Pro) \\
6 & Final assembly: dedup + validation \\
\hline
\end{tabular}
\caption{Evaluation-suite construction phases.}
\label{tab:eval-phases}
\end{table*}

The production run targets 1000 conflict scenarios (100 per pair across 10 pairs), 1000 aligned controls (1:1 with conflicts), and 300 reference baselines (30 per pair). After quality control and deduplication, the final dataset contains 1752 examples.

\subsubsection{Phase 1: scenario generation}
\label{appendix:evaluation-suite:phase1}

Base rows are drawn from Alpaca and Dolly. Before sampling, all rows previously used in SFT or DPO (training or validation) are excluded from the evaluation pool to prevent contamination between training data and the held-out evaluation suite. For each conflict pair, 100 instances are constructed by:
\begin{enumerate}[noitemsep, topsep=0pt]
    \item Sampling a base row from the filtered evaluation pool.
    \item Classifying its domain.
    \item Gathering candidate materials: 2--3 $L_0$ rules from the relevant category, 2--3 domain-matched $L_1$ prompts, 2--3 injection templates (safety-targeting for $L_0$-victim pairs, non-safety otherwise), 2--3 domain-matched $L_4$ candidates, and $L_2$ attribute/value options.
    \item Sending everything to GPT-4o (temperature 0.7) with a structured prompt that names the conflict pair, identifies which level must win and which must lose, and requires that all five levels are topically coherent and that \texttt{evaluation\_criteria} are automatically checkable.
    \item Receiving a structured JSON response with the selected/adapted $L_0$--$L_4$ content, a \texttt{conflict\_description}, \texttt{correct\_behavior}, \texttt{violation\_behavior}, and a list of \texttt{evaluation\_criteria}.
\end{enumerate}
Malformed JSON or validation failures (missing fields, empty criteria) trigger up to two retries with a +0.1 temperature bump per retry; after three failures the scenario is discarded. The final generation stage produces 1000 raw scenarios.

\subsubsection{Phase 2: prompt assembly and gold responses}
\label{appendix:evaluation-suite:phase2}

Each raw scenario is passed through the same prompt-assembly pipeline used for DPO construction, which produces the delimited 5-level prompt and validates delimiter integrity. The gold response is then generated via context distillation using Claude Sonnet 4 (temperature 0.3, retry loop with temperatures [0.3, 0.5, 0.7] on failure) under one of four system-prompt variants: a default variant for most pairs, an $L_4$-attacker variant for the four $L_*$--$L_4$ pairs, an $L_1$-victim variant for $L_0$--$L_1$, and an explicit refusal-expecting variant for all four $L_0$-victim pairs (where the correct response is a polite refusal). Refusal handling distinguishes API-level refusals (which trigger a retry) from expected domain-level refusals (which are accepted). Each scenario is augmented with a gold response and a fully assembled prompt.

\subsubsection{Phase 3: aligned controls}
\label{appendix:evaluation-suite:phase3}

For every conflict scenario we produce one matched control with the conflict removed but everything else held constant. The control allows the over-refusal rate to be measured per pair on prompts whose surface structure is identical to the conflict prompts.

\textbf{Strategy A: replace attacker content ($\sim$800 instances).} For most pairs, the attacking level is overwritten with compatible content via pure string manipulation: in $L_*$--$L_3$ pairs the $L_3$ user message reverts to the original base-row instruction; in $L_*$--$L_4$ pairs the injected $L_4$ is replaced by a benign domain-matched $L_4$ entry (or set to \texttt{null}); in $L_0$--$L_1$ the adversarial $L_1$ is replaced by a benign domain-matched library prompt; in $L_0$--$L_2$ the conflicting attribute is removed.

\textbf{Strategy B: generate compatible replacement ($\sim$200 instances).} For pairs where simple substitution does not produce natural content ($L_1$--$L_2$ format conflicts, cascading-style scenarios), GPT-4o-mini generates a compatible replacement for the attacking level conditioned on the other four levels.

Gold responses for controls are produced by the same Claude Sonnet 4 distillation process as the conflict scenarios but with refusal expectations disabled; when the base-row output is already a natural fit, it is reused directly. Each control carries a link back to its matched conflict scenario and a label identifying whether it was constructed via attacker replacement or model-generated repair.

\subsubsection{Phase 4: reference baselines}
\label{appendix:evaluation-suite:phase4}

Sample 30 conflict scenarios per pair (300 total, uniform random) and strip every architectural cue: remove all special tokens, omit ISE segment IDs, and concatenate the raw level contents into a single newline-separated flat string. All scenario metadata 
carries over, along with a link back to the original conflict instance. The reference split tests how the same model and the same scenario behave when only the architectural signal is removed.

\subsubsection{Phase 5: dual-judge quality control}
\label{appendix:evaluation-suite:phase5}

All 1000 conflict scenarios are dual-judged. Aligned controls and reference baselines are not separately judged because they derive from already-judged conflicts. The judge pair is GPT-4o + Gemini 2.5 Pro, which yields full role independence: GPT-4o generates scenarios in Phase 1, Claude generates gold responses in Phase 2, and GPT-4o + Gemini judge quality in Phase 5. Each judge scores five dimensions on a 1--5 Likert scale: conflict clarity, gold-response correctness, evaluation-criteria checkability, scenario realism, and a ``genuine understanding test'' dimension that asks whether resolving the conflict requires understanding the hierarchy rather than pattern matching. Each judge also returns a binary \texttt{keep} decision.

Decision rules are stricter than for DPO QC: a scenario is \emph{kept} only when both judges set \texttt{keep: true} and all ten scores (five per judge) are at least 4; \emph{discarded and regenerated} when any score is below 3; and \emph{flagged for manual review} when judges disagree on \texttt{keep} or when scores fall in the 3--4 borderline range. Discarded scenarios are replaced once using unused base rows; if a replacement also fails, the slot is dropped (yielding slightly fewer than 100 examples for that pair). 

\subsubsection{Phase 6: final assembly}
\label{appendix:evaluation-suite:phase6}

On passing conflict scenarios we run hash deduplication and embedding-based deduplication at threshold 0.85 (stricter than DPO's 0.92, since evaluation requires maximum scenario diversity). Validation checks confirm delimiter integrity in conflict and aligned splits, absence of delimiters in reference baselines, presence of all required schema fields, that \texttt{evaluation\_criteria} is a non-empty list of two to three entries, and that \texttt{gold\_response} is non-empty and only matches refusal patterns when the victim level is $L_0$.

After quality control and deduplication, the pipeline produces 1752 evaluable instances across the conflict, aligned-control, and reference splits. Aggregate statistics including counts, discard rates, deduplication rates, and judge-score distributions are retained for analysis.

\section{GW-DPO}
\label{appendix:gw-dpo}

This section provides background on standard DPO and ODPO that the main-body presentation of GW-DPO (\S\ref{sec:gravity-weighted-DPO}) presupposes, the inherited ODPO guarantee on the reward gap, and the per-pair margin tables for both schedules.

\subsection{Background: DPO and ODPO}
\label{appendix:dpo-odpo-background}

\textbf{Standard DPO.} \citet{rafailov2023direct} eliminate the explicit reward model of RLHF~\citep{ouyang2022training} by closing a Bradley-Terry preference likelihood~\citep{bradley1952rank} in the policy itself, yielding the loss
\begin{equation}
\begin{aligned}
\mathcal{L}_{\text{DPO}}(\pi_\theta;\pi_{\text{ref}})
&= -\mathbb{E}_{(x,y_w,y_l)\sim\mathcal{D}} \\
&\quad \Big[\log\sigma\big(\beta\cdot\Delta r_\theta(x,y_w,y_l)\big)\Big],
\end{aligned}
\end{equation}
where $\Delta r_\theta$ is the implicit reward margin defined in \S\ref{sec:gravity-weighted-DPO}. The Bradley-Terry derivation guarantees only that $\Delta r_\theta$ becomes positive at the optimum: the chosen response is more likely than the rejected one relative to the reference. It does not specify how large the margin should be, nor does it permit the loss to weight pairs differently. A subtle conflict between adjacent levels and a sharp conflict between the top and bottom levels contribute the same gradient.

\textbf{ODPO.} \citet{amini-etal-2024-direct} address this limitation by introducing a per-sample, non-negative offset $\delta\geq 0$ inside the sigmoid:
\begin{equation}
\small
\begin{aligned}
\mathcal{L}_{\text{ODPO}}(\pi_\theta;\pi_{\text{ref}})
&= -\mathbb{E}_{(x,y_w,y_l)\sim\mathcal{D}} \\
&\quad \Big[\log\sigma\big(\beta\cdot\Delta r_\theta(x,y_w,y_l)-\delta\big)\Big].
\end{aligned}
\end{equation}

The offset functions as a target margin: the loss approaches its minimum only once $\beta\cdot\Delta r_\theta\gg\delta$, that is, once the log-ratio gap satisfies $\Delta r_\theta\geq\delta/\beta$ (formal bound in Appendix~\ref{appendix:inherited-reward-gap-bound}). Setting $\delta=0$ recovers DPO. In their experiments \citet{amini-etal-2024-direct} parameterize the offset as a monotone function of a continuous quality difference between $y_w$ and $y_l$, e.g.\ $\delta=\alpha\cdot\log(r(y_w)-r(y_l))$ when scalar reward scores are available. GW-DPO (\S\ref{sec:gravity-weighted-DPO}) replaces this quality-based parameterization with a structural one keyed on the conflicting hierarchy levels.

\subsection{Inherited Reward-Gap Bound}
\label{appendix:inherited-reward-gap-bound}

GW-DPO inherits its core theoretical guarantee from \citet{amini-etal-2024-direct}. Their Theorem 2 establishes that, when preferences are modeled by a Bradley-Terry distribution with Gumbel noise on the estimated rewards $\hat{r}_\theta(x,y)=\beta\log(\pi_\theta(y\mid x)/\pi_{\text{ref}}(y\mid x))$, the probability that the underlying reward difference exceeds the offset $\delta$ is
\begin{equation}
    \mathbb{P}(R_w-R_l>\delta)\;=\;\sigma\big(\Delta\hat{r}_\theta-\delta\big),
\end{equation}
where $\Delta\hat{r}_\theta=\hat{r}_\theta(x,y_w)-\hat{r}_\theta(x,y_l)=\beta\cdot\Delta r_\theta(x,y_w,y_l)$ is the implicit reward gap. Maximizing this probability via the cross-entropy loss (their Eq.\ 7c) drives $\Delta\hat{r}_\theta\geq\delta$ at the optimum, which corresponds to a policy log-ratio gap of at least $\delta/\beta$ nats. In GW-DPO this bound becomes pair-specific: for every conflict $(L_i,L_j)$ in the dataset $\mathcal{D}_{\text{DPO}}$ the trained model is required to satisfy $\Delta\hat{r}_\theta(x,y_w,y_l)\geq\delta(i,j)/\beta$. Distant or high-privilege conflicts demand larger gaps, adjacent low-privilege conflicts demand smaller gaps, and calibration pairs ($\delta=0$) demand only that the chosen response remain more likely than the rejected one. \citet{amini-etal-2024-direct} also show that ODPO is equivalent to the cost-augmented softmax-margin loss of \citet{gimpel-smith-2010-softmax} when the cost is non-negative, which provides an alternative reading of the offset as a cost assigned to the rejected response. GW-DPO inherits this reading too: $\delta(i,j)$ is the cost the model pays for confusing $L_i$ with $L_j$ in a conflict, and the cost grows with structural severity.

\subsection{Per-Pair Offsets for $k=5$}
\label{appendix:per-pair-offsets}

Instantiating the schedules at $k=5$ produces the per-pair offsets in Table~\ref{tab:gw-dpo-margins}. Calibration entries ($i=j$) are zero under both schedules and are omitted. The bilateral schedule reaches its maximum at the $L_0$-vs-$L_4$ corner ($\delta=16\alpha$), four times the maximum of the linear schedule and sixteen times the minimum non-zero offset, while the linear schedule fans out only between $\alpha$ and $4\alpha$. The most consequential difference is at small gaps with high-privilege victims: an $L_0$-vs-$L_1$ conflict carries weight $4\alpha$ under the bilateral schedule but only $\alpha$ under the linear schedule, capturing the intuition that adjacency at the top of the hierarchy is more dangerous than adjacency at the bottom. Translating to the implicit-reward scale, the required gap for an $L_0$-vs-$L_4$ conflict at $\rho=1$ is 16 nats under the bilateral schedule and 4 nats under the linear schedule, while $L_3$-vs-$L_4$ collapses to 1 nat in both cases.

\begin{table*}[h]
\centering
\begin{tabular}{ccccc}
\hline
Conflict $(L_i,L_j)$ & Gap $j-i$ & Victim weight $k{-}1{-}i$ & $\delta_{\text{bi}}$ & $\delta_{\text{lin}}$ \\
\hline
$L_0$--$L_1$ & 1 & 4 & $4\alpha$  & $\alpha$ \\
$L_0$--$L_2$ & 2 & 4 & $8\alpha$  & $2\alpha$ \\
$L_0$--$L_3$ & 3 & 4 & $12\alpha$ & $3\alpha$ \\
$L_0$--$L_4$ & 4 & 4 & $16\alpha$ & $4\alpha$ \\
$L_1$--$L_2$ & 1 & 3 & $3\alpha$  & $\alpha$ \\
$L_1$--$L_3$ & 2 & 3 & $6\alpha$  & $2\alpha$ \\
$L_1$--$L_4$ & 3 & 3 & $9\alpha$  & $3\alpha$ \\
$L_2$--$L_3$ & 1 & 2 & $2\alpha$  & $\alpha$ \\
$L_2$--$L_4$ & 2 & 2 & $4\alpha$  & $2\alpha$ \\
$L_3$--$L_4$ & 1 & 1 & $\alpha$   & $\alpha$ \\
\hline
\end{tabular}
\caption{Per-pair offsets $\delta(i,j)$ for the bilateral and linear schedules at $k=5$. Calibration entries ($i=j$) are zero under both schedules and are omitted.}
\label{tab:gw-dpo-margins}
\end{table*}

\section{Experiments}
\label{appendix:experiments}

\subsection{Training Hyperparameters}
\label{appendix:training-hyperparameters}

Both phases run on a single NVIDIA A100 80GB GPU in bf16 precision with effective batch size 32 and AdamW~\citep{loshchilov2018decoupled} optimization (cosine LR decay, warmup ratio 0.03, weight decay 0.01). LoRA targets the seven attention and gated-MLP projections (q, k, v, o, gate, up, down) at rank 64, $\alpha=128$, and dropout 0.1; the 12 special delimiter token embeddings and the ISE embedding layer are trained as full-precision parameters alongside the adapters.

SFT (Phase~1) uses a learning rate of 2e-5, maximum sequence length 4096, and three epochs on $\mathcal{D}_{\text{SFT}}$. Loss is computed only on response tokens. DPO (Phase~2) uses a learning rate of 5e-5, maximum sequence length 2048 (DPO p99 $\approx$ 1.3K tokens), and one epoch per curriculum stage. Loss is masked over prompt tokens and computed on the chosen and rejected completions.

\subsection{Curriculum and reference-model updates}
\label{appendix:curriculum-reference-model}

We train the GW-DPO objective in three curriculum stages, ordered by difficulty in the sense of \citet{bengio2009curriculum} and applied to preference optimization in the spirit of Curry-DPO~\citep{pattnaik-etal-2024-enhancing} and Curriculum-DPO++~\citep{croitoru2025curriculum}. Stage 1 (one epoch, easy) restricts the training set to conflicts with privilege gap at least three, namely $(L_0, L_3)$, $(L_0, L_4)$, and $(L_1, L_4)$. 
Stage 2 (one epoch, medium) adds gap-two conflicts $(L_0, L_2)$, $(L_1, L_3)$, and $(L_2, L_4)$. 
Stage 3 (one epoch, hard) adds the four adjacent-level pairs $(L_0, L_1)$, $(L_1, L_2)$, $(L_2, L_3)$, and $(L_3, L_4)$. 
Calibration examples ($\delta = 0$) appear in every stage in proportion to the conflict examples, preventing safety drift from concentrating in the later stages.

The ordering matches the bilateral schedule's severity ranking only approximately: gap-three pairs $(L_0, L_3)$ and $(L_1, L_4)$ both belong to Stage 1 even though they receive different offsets ($12\alpha$ and $9\alpha$ respectively). We retain the gap-based ordering rather than a strict severity-based ordering for two reasons. First, gap is a coarser, more interpretable axis that directly matches the curriculum signal used in Curry-DPO. Second, Stage 1 still strictly precedes Stage 3 in severity under both schedules, which is sufficient for the curriculum structure.

Following sDPO~\citep{kim-etal-2025-sdpo}, we update the reference model between stages rather than holding it frozen for the entire run. The end-of-stage policy of stage $s$ becomes the reference for stage $s+1$. We deviate from sDPO's original prescription in one detail: rather than using the literal end-of-stage checkpoint, we use the in-stage checkpoint with the lowest validation loss, and we re-initialize the policy from this same checkpoint for stage $s+1$. Preliminary runs showed that each single-epoch stage reaches its validation minimum well before the epoch ends and subsequently overfits the cumulative-filtered training subset; anchoring the next stage's KL penalty to an overfit reference compounded this drift across stages. Resetting both the reference and the policy continuation to the best-of-stage snapshot preserves sDPO's progressive-reference property while preventing compounded overfitting.

\subsection{Hyperparameter Tuning}
\label{appendix:hyperparameter-tuning}

\subsubsection{Reparametrization, Search Space, Config Grid}
\label{appendix:reparametrization-search-space-config-grid}

We tune the two hyperparameters $\beta$, the inverse temperature which controls how aggressively the policy $\pi_\theta$ may diverge from $\pi_{\text{ref}}$, and the gravity coefficient $\alpha$, which scales the per-pair structural margin $\delta(i,j)$. Since the implicit-reward target gap satisfies $\Delta r_\theta\geq\delta(i,j)/\beta$, the two interact strongly and $\alpha$ is meaningful only relative to $\beta$; we therefore reparametrize the search in the dimensionless ratio $\rho=\alpha/\beta$. 

SecAlign~\citep{chen2025secalign} fixes $\beta=0.1$ and the community DPO range is $\{0.05,0.1,0.2\}$; $\alpha$ has no precedent. All other DPO hyperparameters (learning rate, scheduler, warmup, weight decay, LoRA configuration, batch size) are adopted from SecAlign. 

Sweeping $(\alpha,\beta)$ independently would cover the $(\rho,\beta)$ plane unevenly and double-count the regime where the two coefficients move in lockstep, so we sweep $(\rho,\beta)$ and derive $\alpha$. Well-converged DPO models produce per-pair reward gaps in the range of one to five nats, which bounds the realistic sweep range; targets above ten nats are aspirational and may simply fail to saturate without affecting the per-pair sign of $\Delta r_\theta$ (which is what the model is ultimately ranked on). The cell $\rho=0$ recovers standard DPO without the gravity term and is intentionally excluded from the sweep.

The full grid is $\rho\in\{0.5,1,2,3\}\times\beta\in\{0.05,0.1,0.2\}$ (twelve cells). Because a full twelve-cell sweep under the three-stage curriculum is prohibitively expensive, we run a two-phase axial design over six cells. Phase~1 fixes $\rho=1$ (so the per-pair margin is $\beta\cdot(j-i)$ and the loss target $\Delta r_\theta$ is independent of $\beta$, isolating the $\beta$ axis) and varies $\beta\in\{0.05,0.1,0.2\}$. Phase~2 fixes the phase-1 winner $\beta=0.1$ and varies $\alpha\in\{0.05,0.20,0.30\}$, equivalently $\rho\in\{0.5,2,3\}$; the phase-1 winner is reused as the $\rho=1$ anchor so phase~2 covers the four $\rho$ values without a redundant re-run. The six cells executed and their target reward gaps at the maximal hierarchy gap of four are depicted in Table~\ref{tab:search-grid}. 

\begin{table*}[t]
\centering
\begin{tabular}{c c c c c}
\hline
Phase & $\rho$ & $\beta$ & $\alpha$ & Target $\Delta r$ at gap=4 \\
\hline
1 & 1.0 & 0.05 & 0.05 & 4 nats \\
1 & 1.0 & 0.10 & 0.10 & 4 nats \\
1 & 1.0 & 0.20 & 0.20 & 4 nats \\
2 & 0.5 & 0.10 & 0.05 & 2 nats \\
2 & 2.0 & 0.10 & 0.20 & 8 nats \\
2 & 3.0 & 0.10 & 0.30 & 12 nats \\
\hline
\end{tabular}
\caption{Six configurations from a two-phase axial search over $(\rho,\beta)$, showing target reward gaps $\Delta r_\theta$ at hierarchy gap 4.}
\label{tab:search-grid}
\end{table*}

\subsubsection{Held-Out HP-Select Split}
\label{appendix:held-out-hp-split}

The full DPO validation set has 2573 preference pairs. To avoid leaking ranking signals, we hold out a stratified subset of 1000 pairs instead of using the full set for both training-time evaluation and post-training ranking. Stratification is by (\text{level\_gap}, \text{is\_calibration}), with proportional allocation and largest-remainder rounding. Calibration pairs all have gap 0, making it the only multi-bucket gap-0 stratum. The held-out set contains 703 calibration pairs, 85 gap-1, 101 gap-2, 81 gap-3, and 30 gap-4; the remaining 1,573 pairs are used for training-time validation.

\subsubsection{HP Training and Fixed Hyperparameters}
\label{appendix:hp-training}

Each configuration runs the same DPO pipeline as the production training script, with two compute-saving simplifications. First, curriculum training is disabled: each configuration runs a single unfiltered stage on the full $\sim$8.7K-pair training set instead of the three-stage easy-to-hard curriculum, reducing compute time. The trade-off is that the curriculum's sDPO~\citep{kim-etal-2025-sdpo} reference update between stages is not exercised; this is acceptable because the ranking signal is dominated by the per-pair reward margin rather than by the inter-stage dynamics. Second, the final LoRA adapter is not merged into a standalone model after each cell; only the per-config adapter and ISE weights are persisted. The winner is re-merged once after the sweep, in the production training run.

Training settings held fixed across the sweep are inherited from the base configuration and match the SecAlign DPO recipe~\citep{chen2025secalign}: AdamW~\citep{loshchilov2018decoupled} with learning rate 5e-5, cosine LR schedule with warmup ratio 0.03, weight decay 0.01, one epoch per stage (single stage in HP search), effective batch size of 32, maximum sequence length 2048 (DPO p99 $\approx$ 1.3K tokens), bf16 precision, and the metric for best model is eval loss (lower is better). LoRA uses rank 64, $\alpha=128$, dropout 0.1, and targets the seven attention and gated-MLP projections (q, k, v, o, gate, up, down). The margin schedule used in the sweep is the linear schedule $\delta=\alpha\cdot(j-i)$ with calibration rows forced to zero. 

\subsubsection{HP Selection Metric}
\label{appendix:hp-selection-metric}

For each configuration, after training, the policy and reference are run forward over the 1000-pair held-out cut. Per-pair log-probabilities are computed with masking over the completion tokens. Per-pair rewards are $r_w=\beta\cdot(\log\pi_\theta(y_w)-\log\pi_{\text{ref}}(y_w))$ and $r_l=\beta\cdot(\log\pi_\theta(y_l)-\log\pi_{\text{ref}}(y_l))$; a pair is correct iff $r_w>r_l$ strictly. Per-pair accuracy is invariant to the $\beta$ scalar, so using each configuration's own $\beta$ only affects the reported mean reward margin, not the ranking. Three aggregated metrics are reported. The \textit{per-gap accuracy} is computed for each $\text{gap}\in\{0,1,2,3,4\}$. The \textit{macro-average accuracy} is the per-gap accuracy averaged with equal weight per populated gap bucket; it is the primary ranking metric and stops the gap-zero calibration bucket from dominating. The \textit{gap-weighted accuracy} is $\sum_g g\cdot(\text{correct}_g)/\sum_g g\cdot n_g$, used as a tie-breaker to emphasize high-gap performance, mirroring the gravity-weighting ``philosophy'' of the loss. Mean reward margin is reported as a diagnostic only. The eval loss on the training-time eval set is \textit{not} used for ranking, since its scale depends on $\alpha$ and thus cross-config comparison is not feasible.

\subsubsection{HP Search Results}
\label{appendix:hp-search-results}

\paragraph{Phase 1: $\beta$ sweep at $\rho=1$.} Three runs at $\beta\in\{0.05,0.1,0.2\}$ with $\alpha=\beta$, ranked by macro-average reward accuracy on the 1000-pair held-out dataset are depicted in Table~\ref{tab:hp-phase1-results}. 

\begin{table*}[t]
\centering
\begin{tabular}{c c c c c c c c c c}
\hline
Rank & $\beta$ & $\alpha$ & macro-avg & gap-wtd & gap0 & gap1 & gap2 & gap3 & gap4 \\
\hline
1 & 0.20 & 0.20 & 0.9927 & 0.9954 & 0.999 & 0.965 & 1.000 & 1.000 & 1.000 \\
2 & 0.10 & 0.10 & 0.9923 & 0.9923 & 0.997 & 0.976 & 1.000 & 0.988 & 1.000 \\
3 & 0.05 & 0.05 & 0.9862 & 0.9846 & 0.999 & 0.965 & 0.980 & 0.988 & 1.000 \\
\hline
\end{tabular}
\caption{Phase 1 results: sweep over $\beta \in \{0.05, 0.1, 0.2\}$ with $\alpha=\beta$ at $\rho=1$, ranked by macro-average reward accuracy on a 1{,}000-pair held-out set.}
\label{tab:hp-phase1-results}
\end{table*}

The $\beta=0.2$ and $\beta=0.1$ runs are within $0.0004$ macro-average of each other, well inside the noise floor implied by the 30-record gap-4 bucket. We select $\beta=0.1$ as the phase-1 winner: it matches the SecAlign~\citep{chen2025secalign} default, produces the higher gap-1 accuracy ($0.976$ versus $0.965$, where gap-1 is the discriminating bucket), and pins the KL ``leash'' at the standard DPO temperature for the phase-2 $\alpha$ sweep.

\paragraph{Phase 2: $\alpha$ sweep at $\beta=0.1$.} Three new runs at $\alpha\in\{0.05,0.20,0.30\}$ (equivalently $\rho\in\{0.5,2,3\}$) plus the phase-1 winner as the $\rho=1$ anchor are depicted in Table~\ref{tab:hp-phase2-results}. 

\begin{table*}[t]
\centering
\begin{tabular}{c c c c c c c c c c}
\hline
Rank & $\rho$ & $\alpha$ & macro-avg & gap-wtd & gap0 & gap1 & gap2 & gap3 & gap4 \\
\hline
1 & 3.0 & 0.30 & 0.9965 & 0.9985 & 0.994 & 0.988 & 1.000 & 1.000 & 1.000 \\
2 & 2.0 & 0.20 & 0.9953 & 0.9969 & 1.000 & 0.976 & 1.000 & 1.000 & 1.000 \\
3 & 1.0 & 0.10 & 0.9923 & 0.9923 & 0.997 & 0.976 & 1.000 & 0.988 & 1.000 \\
4 & 0.5 & 0.05 & 0.9885 & 0.9877 & 1.000 & 0.965 & 0.990 & 0.988 & 1.000 \\
\hline
\end{tabular}
\caption{Phase 2 results: sweep over $\alpha \in \{0.05, 0.20, 0.30\}$ at fixed $\beta=0.1$ (varying $\rho$), including the $\rho=1$ anchor from Phase 1, ranked by macro-average reward accuracy on the held-out set.}
\label{tab:hp-phase2-results}
\end{table*}

The macro-average is increasing in $\rho$ over the swept range; the config at $\rho=3$, $\beta=0.1$, $\alpha=0.3$ is the overall winner. Three cross-phase observations support stopping at $\rho=3$. First, all six runs solve gap four perfectly on the 30-pair bucket; the bucket cannot discriminate at this performance level, since the $\pm 9$pp margin of error is wider than the entire spread across the table. Second, gap one is the discriminating bucket: adjacent-level pairs are the hardest, with the winner reaching $0.988$ and the worst run reaching $0.965$. Third, $\rho$ is the dominant axis: holding $\beta=0.1$, the macro average traces $0.9885\to0.9923\to0.9953\to0.9965$ as $\rho$ moves through $\{0.5,1,2,3\}$. The monotone trend motivated stopping at $\rho=3$ on the assumption that gap-1 is saturating and further gravity simply makes the loss target unreachable without affecting the per-pair sign of $\Delta r_\theta$; verifying this would require runs at $\rho\geq4$ which were not executed.

\subsection{Reference and Ablation Studies}
\label{appendix:reference-ablation-studies}

This appendix records the materialization of each of the five non-GW-DPO configurations evaluated in \S~\ref{sec:reference-ablations}. The two GW-DPO variants (GW-DPO~(lin.) and GW-DPO~(bi.)) follow the production training pipeline of \S~\ref{sec:training} with the linear and bilateral schedules from \S~\ref{sec:gravity-weighted-DPO}, the hyperparameters $\rho=3$, $\beta=0.1$, $\alpha=0.3$ from the sweep (\S~\ref{sec:hyperparameter-tuning}), and the three-stage curriculum with sDPO reference updates~\citep{kim-etal-2025-sdpo,pattnaik-etal-2024-enhancing}. Each subsection below lists the construction recipe of one ablation, the hyperparameters held constant relative to GW-DPO~(lin.), and the design choices that prevent confounds.

\subsubsection{\textit{baseline}}
\label{appendix:ablations-baseline}

The \textit{baseline} configuration is Llama-3.1-8B-Instruct~\citep{grattafiori2024llama3herdmodels} with the 12 hierarchy delimiter tokens (\texttt{<|L0\_START|>}, \texttt{<|L0\_END|>}, \dots, \texttt{<|L4\_START|>}, \texttt{<|L4\_END|>}) and the two response delimiters added to the tokenizer vocabulary, the embedding matrix resized accordingly, and the new rows mean-initialized from the existing token embeddings exactly as in the SFT setup (\S~\ref{sec:architecture}). No training is applied, no Instructional Segment Embedding layer is attached, and no LoRA adapters are loaded. \textit{baseline} shares its tokenizer and embedding architecture with all DPO configurations, so the \textit{baseline}\,$\to$\,\textit{SFT} gap measures the effect of training itself rather than the joint effect of training and tokenizer extension.

We deliberately do not use the off-the-shelf checkpoint with its original tokenizer as the ablation floor. Under the original tokenizer the delimiter strings are split into BPE subwords (\texttt{<}, \texttt{|}, \texttt{L0\_START}, \texttt{|}, \texttt{>}), so an off-the-shelf row would conflate two distinct effects: extending the tokenizer with hierarchy delimiters, and training on hierarchy data via SFT. A small-scale probe of the genuinely off-the-shelf model on delimited prompts returned the empty string on the majority of conflict and aligned-control instances; the same model handled flat-text reference prompts without issue. The probe therefore measures whether the model can parse a delimited prompt at all rather than whether it can comply with a hierarchy. We report this finding in passing here and use \textit{baseline} as the lower bound in the headline ablation table.

\subsubsection{\textit{SFT}}
\label{appendix:ablations-sft}

The \textit{SFT} configuration is the Phase~1 SFT checkpoint (\S~\ref{sec:training-sft}) with the LoRA adapter merged into the base via \texttt{PeftModel.merge\_and\_unload()}, the per-row deltas for the 12 trainable special-token embeddings folded in, and the trained ISE weights co-located alongside the merged model directory. The evaluation loader auto-detects the ISE weights and wraps the model in the ISE-aware variant of the architecture (\S~\ref{sec:architecture}), so \textit{SFT} shares its generation code path with the DPO configurations. This is the same merge that the GW-DPO trainer performs internally between Phase~1 and Phase~2. The \textit{baseline}\,$\to$\,\textit{SFT} gap therefore measures the effect of SFT training (LoRA, ISE, and trained delimiter rows) holding the tokenizer architecture constant; the \textit{SFT}\,$\to$\,\textit{DPO}\,$\to$\,GW-DPO~(lin.)\,$\to$\,GW-DPO~(bi.) gaps measure the marginal effect of standard DPO, the linear gravity schedule, and the bilateral gravity schedule respectively.

\subsubsection{\textit{DPO}}
\label{appendix:ablations-dpo}

The \textit{DPO} configuration runs the same training pipeline as the GW-DPO models with one parameter changed: the gravity coefficient is set to $\alpha=0$. The per-pair offset $\delta(i,j)=\alpha\cdot(j-i)$ collapses to zero for every conflict, so the GW-DPO loss in \S~\ref{sec:gravity-weighted-DPO} reduces algebraically to standard DPO~\citep{rafailov2023direct}, $-\log\sigma(\beta\cdot\Delta r_\theta)$. All other hyperparameters are held at their GW-DPO~(lin.) values: $\beta=0.1$, the three-stage curriculum with sDPO reference updates, the same LoRA configuration, and the same SFT-merged initialization. We do not re-tune $\beta$: re-tuning would conflate two effects --- removing gravity weighting and shifting the DPO temperature --- whereas the question asked of \textit{DPO}\,$\to$\,GW-DPO~(lin.) is the marginal effect of gravity weighting alone. Holding the curriculum on for \textit{DPO} is similarly a design choice: disabling it would conflate ``no gravity'' with ``no curriculum''.

\subsubsection{\textit{No-ISE}}
\label{appendix:ablations-no-ise}

The \textit{No-ISE} configuration removes the ISE layer while retaining the 12 special delimiter tokens and the entire SFT-plus-GW-DPO~(lin.) training pipeline. The architecture reduces to the special-tokens-only setup of StruQ~\citep{chen2025struq}, but trained with our five-level data, our gravity-weighted loss, and our curriculum. All other hyperparameters are held at their GW-DPO~(lin.) values. The GW-DPO~(lin.)\,$\to$\,\textit{No-ISE} gap therefore isolates the marginal contribution of the per-token segment signal beyond what the discrete delimiter boundaries already encode. This is the configuration that grounds the discussion in \S~\ref{sec:discussion}: removing ISE changes conflict accuracy (macro PPA) by less than 0.05 but shifts the over-refusal rate from 0.024 to 0.576, consistent with the mechanistic prediction~\citep{arditi2024refusal,qi2024safety} that a small per-token signal can shape refusal behavior without altering the underlying conflict reasoning.

\subsubsection{\textit{3-level~(lin.)}}
\label{appendix:ablations-3-level}

The \textit{3-level~(lin.)} configuration is the comparison against the prior state of the art~\citep{wallace2024instruction}. It collapses the five-level training data into a Wallace-style three-tier hierarchy at training time: the $L_0$, $L_1$, and $L_2$ spans are concatenated and rendered inside the existing \texttt{<|L0\_START|>...<|L0\_END|>} wrapper as a single ``System'' tier, $L_3$ is preserved as ``User'', and $L_4$ is preserved as ``Tool''. Architecture, tokenizer, the ISE layer, and all DPO hyperparameters ($\beta=0.1$, $\alpha=0.3$, LoRA, sDPO) are held at their GW-DPO~(lin.) values, so the model has five-level capacity but trains on three-level content. The GW-DPO~(lin.)\,$\to$\,\textit{3-level~(lin.)} gap therefore attributes cleanly to the depth of the training hierarchy, with architecture and tuning held constant.

Two adaptations are required by the depth collapse. First, intra-System training pairs ($L_0$-vs-$L_1$, $L_0$-vs-$L_2$, $L_1$-vs-$L_2$, approximately 17\% of the train set and 6\% of the validation set) are dropped because the three-level hierarchy has no privilege axis to resolve them. Second, the curriculum compresses to two stages because the maximum hierarchy gap under the collapsed structure is two: stage~1 keeps only gap-2 pairs (System-vs-Tool), and stage~2 adds the two gap-1 pairs. The same cumulative filtering, calibration handling, and best-of-stage sDPO logic apply.

At evaluation time the same collapse function rewrites every conflict and aligned-control prompt in-memory before tokenization; the reference-baseline split is unaffected because flat-text prompts contain no delimiters. PPA on the intra-System conflict scenarios is expected to collapse to floor for \textit{3-level~(lin.)} because the three-level hierarchy cannot distinguish levels within the System block, so macro PPA and the gap-weighted aggregate are expected to drop relative to GW-DPO~(lin.) primarily through that structural penalty rather than through poor performance on representable pairs. This is the empirical basis for the verdict in \S~\ref{sec:discussion} that five-level training preserves the broader resolution capability while three-level training delivers narrower mastery of the System-User-Tool subset.

\subsection{Evaluation Metrics}
\label{appendix:evaluation-metrics}

This appendix provides the mathematical definitions and aggregation rules underlying the three metric families summarized in \S\ref{sec:evaluation-metrics}, the per-record judge schema from which all aggregates are derived, and the response-shape diagnostics that separate ``did the model emit text at all?'' from ``given that it did, did it pick the right level?''. Throughout, $L_i$ denotes hierarchy level $i\in\{0,\dots,k-1\}$ with $k=5$, and a conflict pair $(L_i,L_j)$ with $i<j$ has \textit{victim} $L_i$ (higher priority) and \textit{attacker} $L_j$ (lower priority). The conflict, reference, and aligned-control splits are as defined in \S\ref{sec:evaluation-suite}.

\subsubsection{Hierarchy Resolution}
\label{appendix:metrics-resolution}

\paragraph{Pairwise Priority Adherence (PPA).} Out of all cases where $L_i$ and $L_j$ conflict, how often does the model correctly prioritize the higher one without mistakenly following the lower one? For each conflict pair $(L_i,L_j)$, let $\mathcal{S}_{i,j}$ denote the set of judged conflict-split records belonging to that pair, with $N_{i,j}=|\mathcal{S}_{i,j}|$. The judge returns per-record booleans $\mathrm{sat}_r$ (\textit{satisfies higher}) and $\mathrm{foll}_r$ (\textit{follows lower}); a record is PPA-correct iff both conditions hold jointly:
\begin{equation}
\mathrm{PPA}_{i,j}\;=\;\frac{1}{N_{i,j}}\sum_{r\in\mathcal{S}_{i,j}}\mathbb{1}\!\left[\mathrm{sat}_r\wedge\neg\mathrm{foll}_r\right].
\end{equation}
Records missing either field (e.g.\ judge parse failures) are counted as PPA-incorrect, so the metric fails closed. The conjunction is required because ``satisfies higher \textit{and} follows lower'' is internally inconsistent and almost always picks the wrong action when forced to commit; a pure ``satisfies higher'' criterion would credit such responses, and a pure ``did not follow lower'' would credit empty or off-topic outputs. The macro-average over the set $\mathcal{P}=\{(i,j):N_{i,j}>0\}$ of populated pairs is
\begin{equation}
\mathrm{PPA}_{\mathrm{macro}}\;=\;\frac{1}{|\mathcal{P}|}\sum_{(i,j)\in\mathcal{P}}\mathrm{PPA}_{i,j},
\end{equation}
matching the equal-weight aggregation used by prior hierarchy-compliance papers and reported alongside the gap-weighted score below for cross-paper comparability.

\paragraph{Weighted Hierarchy Score (WHS).} A version of macro PPA that gives more weight to wide-gap conflicts, on the grounds that overriding a platform rule with a tool output is structurally more dangerous than overriding a user message. Let $g_{i,j}=j-i\in\{1,\dots,k-1\}$ denote the privilege gap of pair $(L_i,L_j)$. The gap-weighted aggregate is
\begin{equation}
\mathrm{WHS}\;=\;\frac{\sum_{(i,j)\in\mathcal{P}}g_{i,j}\cdot\mathrm{PPA}_{i,j}}{\sum_{(i,j)\in\mathcal{P}}g_{i,j}}.
\end{equation}
WHS encodes the operational asymmetry that long-range violations are more severe than adjacent-level ones (a tool output overriding a platform rule is structurally more dangerous than a tool output overriding a user message), in the same spirit as the gravity-weighted offset of \S\ref{sec:gravity-weighted-DPO}. We additionally report per-gap PPA averages within each bucket $g\in\{1,\dots,k-1\}$,
\begin{align}
\mathrm{PPA}_{g} &= \frac{1}{|\mathcal{P}_g|}\sum_{(i,j)\in\mathcal{P}_g}\mathrm{PPA}_{i,j}, \\
\mathcal{P}_g &= \{(i,j)\in\mathcal{P}:j-i=g\},
\end{align}
as a diagnostic: hierarchy reasoning should improve monotonically with $g$, since wider asymmetries are easier to resolve.

\paragraph{Attack Success Rate (ASR).} How often does the attacker level win the conflict? The complement of PPA, reported in the convention adversarial-robustness readers expect. For comparability with the prompt-injection and jailbreak literature we report the complementary rate
\begin{equation}
\mathrm{ASR}_{i,j}\;=\;1-\mathrm{PPA}_{i,j}.
\end{equation}
By construction $\mathrm{ASR}_{i,j}+\mathrm{PPA}_{i,j}=1$, so ASR carries no additional information beyond PPA; it is reported as a re-presentation in the convention attack-literature readers expect.

\subsubsection{Generalization}
\label{appendix:metrics-generalization}

\paragraph{Reference PPA.} The same hierarchy-resolution question as PPA, but asked on flat-text rewrites of the same scenarios with all level wrappers removed. Formally, the same PPA definition is applied to the \textit{reference} split, where each conflict scenario is rewritten as a single concatenated natural-language prompt without the level-token wrappers (\S\ref{sec:evaluation-suite}). Denoting reference-split records $\mathcal{S}^{\mathrm{ref}}_{i,j}$,
\begin{equation}
\mathrm{PPA}^{\mathrm{ref}}_{i,j}\;=\;\frac{1}{N^{\mathrm{ref}}_{i,j}}\sum_{r\in\mathcal{S}^{\mathrm{ref}}_{i,j}}\mathbb{1}\!\left[\mathrm{sat}_r\wedge\neg\mathrm{foll}_r\right],
\end{equation}
with macro-average $\mathrm{PPA}^{\mathrm{ref}}_{\mathrm{macro}}$ defined analogously. A model that has overspecialized to the delimited format scores well on the conflict split but poorly on the reference split; a model that has internalized the underlying priority semantics scores similarly on both.

\paragraph{Utility Delta.} Per-pair gap between the model's performance on the wrapped and flat-text formats of the same scenario: positive means the wrappers help, negative means they hurt. Following IHEval~\citep{zhang-etal-2025-iheval}, we report the per-pair signed gap between conflict-split and reference-split PPA:
\begin{equation}
\Delta_{i,j}\;=\;\mathrm{PPA}_{i,j}-\mathrm{PPA}^{\mathrm{ref}}_{i,j}.
\end{equation}
Aggregated over the set $\mathcal{C}$ of pairs populated in both splits, the signed mean and the mean-absolute (volatility) are
\begin{align}
\bar{\Delta} &= \frac{1}{|\mathcal{C}|}\sum_{(i,j)\in\mathcal{C}}\Delta_{i,j}, \\
\bar{\Delta}_{\mathrm{abs}} &= \frac{1}{|\mathcal{C}|}\sum_{(i,j)\in\mathcal{C}}|\Delta_{i,j}|.
\end{align}
$\Delta_{i,j}<0$ indicates that the architectural cues are costing utility relative to the flat-text equivalent; $\Delta_{i,j}\approx 0$ is the desired regime in which hierarchy reasoning has not reduced general task ability; $\Delta_{i,j}>0$ indicates that the level tokens are actively helping the model resolve the conflict beyond what flat semantics provide. The signed mean and the mean-absolute serve different diagnostic purposes: a near-zero $\bar{\Delta}$ paired with small per-pair variance is a clean alignment between formats, whereas a near-zero $\bar{\Delta}$ paired with large $\bar{\Delta}_{\mathrm{abs}}$ indicates that positives and negatives cancel and per-pair behavior is uneven. Pairs missing from either split are excluded from $\mathcal{C}$ to prevent unpopulated pairs being treated as a baseline of zero.

\subsubsection{Safety Calibration}
\label{appendix:metrics-safety}

\paragraph{Over-Refusal Rate (ORR).} How often does the model wrongly refuse a benign prompt that was paired one-to-one with a conflict scenario as its non-adversarial control? Each conflict scenario in the evaluation suite is paired one-to-one with an aligned-control scenario via a shared \texttt{matched\_conflict\_id} (\S\ref{sec:evaluation-suite}). The aligned controls are benign prompts that the model should answer; refusing one signals over-cautious behavior induced by hierarchy training. Let $\mathcal{A}$ denote the set of aligned-control records whose \texttt{matched\_conflict\_id} resolves to a known conflict pair, and let $\mathrm{ref}_r\in\{0,1\}$ denote the refusal flag produced by the two-stage classifier described below. The aggregate and per-pair rates are
\begin{align}
\mathrm{ORR} &= \frac{1}{|\mathcal{A}|}\sum_{r\in\mathcal{A}}\mathrm{ref}_r, \\
\mathrm{ORR}_{i,j} &= \frac{1}{|\mathcal{A}_{i,j}|}\sum_{r\in\mathcal{A}_{i,j}}\mathrm{ref}_r,
\end{align}
where $\mathcal{A}_{i,j}\subseteq\mathcal{A}$ is the subset whose matched conflict belongs to pair $(L_i,L_j)$. ORR is bounded below by the rate at which benign prompts genuinely trigger refusal-shaped tokens that the regex classifier matches even on a safe response (around $0.04$ for the untrained Llama-3.1-8B-Instruct baseline).

\paragraph{Two-stage refusal classification.} The refusal flag $\mathrm{ref}_r$ is produced by a regex-then-judge cascade. Stage A matches the lower-cased response against a curated list of standard AI refusal phrases (e.g.\ \textit{``I cannot help''}, \textit{``I'm unable to''}, \textit{``as an AI language model''}); a non-match returns $\mathrm{ref}_r=0$ immediately. Stage B is invoked only when the regex matches and the response is long enough that a hedge-then-comply pattern is plausible (length $\geq 600$ characters by default): the response is sent to a GPT-4o judge that returns \texttt{true} iff the model refused all the way through, and \texttt{false} when an apologetic preamble is followed by substantive task help. Short regex-matched responses are conclusive refusals. The same flag is reported as a diagnostic on the conflict and reference splits to expose the refusal disposition across all splits, not just on safe prompts: a high conflict-split refusal rate paired with a low aligned-split rate is the desirable pattern (refuses adversarial prompts, complies with safe ones); the inverse is the failure pattern of an over-cautious model.

\subsubsection{Per-Record Judge Schema}
\label{appendix:metrics-judge}

All boolean fields underlying the aggregates above are produced by a GPT-4o PPA judge that observes the prompt, the model response, and the scenario metadata constructed in \S\ref{sec:evaluation-suite} (victim level, attacker level, conflict description, correct-behavior and violation-behavior statements, and the machine-checkable evaluation criteria). The judge returns strict JSON of the form

\begin{lstlisting}
{
  "satisfies_higher_level": bool,
  "follows_lower_level": bool,
  "refuses": bool,
  "per_criterion": bool[n_{i,j}],
  "score": 1..5,
  "reasoning": string
}
\end{lstlisting}
where \texttt{bool} is either 0 or 1 and $n_{i,j}$ is the number of evaluation criteria attached to scenario $(i,j)$. The two boolean flags drive PPA; the integer Likert score
\begin{equation}
\bar{s}\;=\;\frac{1}{|\mathcal{S}|}\sum_{r\in\mathcal{S}}s_r,\qquad s_r\in\{1,\dots,5\},
\end{equation}
captures partial credit that the boolean fields cannot represent (e.g.\ responses that identify the conflict but only half-resolve it score $3$ or $4$ rather than $5$ or $1$). 

\subsubsection{Response-Shape Diagnostics and Non-Empty Decomposition}
\label{appendix:metrics-decomposition}

The trained model emits \texttt{<|RESP\_END|>} at the end of its response; the evaluation script truncates everything from this token onward at decode time. When an under-trained model emits \texttt{<|RESP\_END|>} as one of its first generated tokens, the response is truncated to the empty string, and the headline PPA bundles two distinct failure modes (``did not answer'' and ``answered but picked the wrong level'') into a single number. To separate them, we report a per-split completion rate
\begin{equation}
\begin{split}
\mathrm{completion}(\mathcal{R})
= 1-\frac{1}{|\mathcal{R}|}
\sum_{r\in\mathcal{R}}
\\
\mathbb{1}\big[
\mathrm{strip}(r.\mathrm{response})
=\varnothing
\big],
\end{split}
\end{equation}
where $\mathcal{R}$ is the set of response records on a given split, and a non-empty variant of every aggregate metric in \S\ref{appendix:metrics-resolution}--\S\ref{appendix:metrics-generalization} that excludes empty responses before aggregation. The non-empty PPA tells us \textit{given that the model answered, did it pick the right level?}; the completion rate tells us \textit{did the model answer at all?}. The two factors decompose the headline number multiplicatively per pair,
\begin{equation}
\mathrm{PPA}^{\mathrm{full}}_{i,j}\;\approx\;\mathrm{completion}_{i,j}\cdot\mathrm{PPA}^{\mathrm{non\text{-}empty}}_{i,j},
\end{equation}
which is exact at the per-pair level and approximate at the macro level. The non-empty utility delta is computed only over pairs populated in both splits' non-empty subsets, so it does not pick up an artefactual signal from one split having more empty rows than the other.

\subsection{Evaluation Suite Results}
\label{appendix:eval-results}

This appendix reports the full per-pair tables underlying the headline aggregates of \S\ref{sec:5-level-evaluation-suite}, together with the response-shape and judge-decomposition diagnostics that those aggregates summarize. The notation follows \S\ref{appendix:evaluation-metrics}: $L_x$\,vs.\,$L_y$ denotes a conflict pair with victim $L_x$ ($x<y$) and attacker $L_y$, and metrics are conditioned on the pair's pair-specific record set (sizes 33--68 per pair on the conflict split, $\sim$30 on the reference split, all 994 aligned-control records on the ORR breakdown).

\subsubsection{Per-Pair PPA on the Conflict and Reference Splits}
\label{appendix:eval-perpair}

Table~\ref{tab:eval-perpair-conflict} reports $\mathrm{PPA}_{i,j}$ for all ten pairs on the conflict split. Table~\ref{tab:eval-perpair-reference} reports $\mathrm{PPA}^{\mathrm{ref}}_{i,j}$ on the reference split. Table~\ref{tab:eval-perpair-asr} re-presents the conflict-split numbers as $\mathrm{ASR}_{i,j}=1-\mathrm{PPA}_{i,j}$ for comparability with the prompt-injection literature.

\begin{table*}[t]
\centering
\small
\begin{tabular}{l c c c c c c c}
\hline
Pair & baseline & SFT & GW-DPO (bi.) & GW-DPO (lin.) & DPO & No-ISE & 3-level~(lin.) \\
\hline
$L_0$\,vs.\,$L_1$ & 0.222 & 0.333 & 0.722 & 0.667 & 0.778 & 0.694 & \textbf{0.750} \\
$L_0$\,vs.\,$L_2$ & 0.303 & 0.303 & 0.879 & 0.848 & \textbf{0.909} & 0.848 & 0.818 \\
$L_0$\,vs.\,$L_3$ & 0.390 & 0.463 & \textbf{0.976} & 0.927 & 0.951 & 0.902 & 0.951 \\
$L_0$\,vs.\,$L_4$ & 0.350 & 0.575 & \textbf{0.975} & 0.750 & 0.900 & 0.875 & \textbf{0.975} \\
$L_1$\,vs.\,$L_2$ & 0.200 & 0.089 & \textbf{0.778} & 0.489 & 0.622 & 0.556 & 0.511 \\
$L_1$\,vs.\,$L_3$ & 0.162 & 0.250 & \textbf{0.926} & 0.735 & \textbf{0.926} & \textbf{0.926} & 0.706 \\
$L_1$\,vs.\,$L_4$ & 0.231 & 0.404 & \textbf{0.904} & 0.692 & 0.827 & 0.596 & 0.788 \\
$L_2$\,vs.\,$L_3$ & 0.170 & 0.113 & 0.642 & 0.660 & 0.679 & \textbf{0.736} & 0.585 \\
$L_2$\,vs.\,$L_4$ & 0.216 & 0.432 & \textbf{0.838} & 0.757 & 0.811 & 0.405 & 0.811 \\
$L_3$\,vs.\,$L_4$ & 0.208 & 0.358 & 0.736 & 0.660 & 0.736 & 0.226 & \textbf{0.887} \\
\hline
\end{tabular}
\caption{Per-pair PPA on the conflict split. Best per row in bold. Pair sizes range 33--68; macro and per-gap aggregates are reported in \S\ref{sec:5-level-evaluation-suite}.}
\label{tab:eval-perpair-conflict}
\end{table*}

\begin{table*}[t]
\centering
\small
\begin{tabular}{l c c c c c c c}
\hline
Pair & baseline & SFT & GW-DPO (bi.) & GW-DPO (lin.) & DPO & No-ISE & 3-level~(lin.) \\
\hline
$L_0$\,vs.\,$L_1$ & 0.767 & 0.667 & 0.833 & 0.900 & 0.800 & \textbf{0.933} & 0.767 \\
$L_0$\,vs.\,$L_2$ & 0.700 & 0.667 & \textbf{0.967} & \textbf{0.967} & 0.833 & 0.900 & 0.933 \\
$L_0$\,vs.\,$L_3$ & 0.867 & 0.733 & 0.967 & \textbf{1.000} & 0.967 & \textbf{1.000} & 0.967 \\
$L_0$\,vs.\,$L_4$ & 0.633 & 0.667 & \textbf{0.900} & \textbf{0.900} & \textbf{0.900} & 0.867 & 0.833 \\
$L_1$\,vs.\,$L_2$ & 0.633 & 0.300 & 0.700 & \textbf{0.767} & 0.667 & \textbf{0.767} & 0.567 \\
$L_1$\,vs.\,$L_3$ & 0.700 & 0.433 & 0.867 & \textbf{0.967} & 0.833 & 0.933 & 0.767 \\
$L_1$\,vs.\,$L_4$ & 0.600 & 0.567 & \textbf{0.867} & 0.833 & \textbf{0.867} & 0.833 & 0.833 \\
$L_2$\,vs.\,$L_3$ & 0.300 & 0.167 & 0.500 & 0.567 & 0.533 & \textbf{0.667} & 0.433 \\
$L_2$\,vs.\,$L_4$ & 0.767 & 0.367 & 0.667 & 0.833 & 0.833 & 0.733 & \textbf{0.933} \\
$L_3$\,vs.\,$L_4$ & 0.633 & 0.367 & 0.667 & 0.700 & 0.667 & 0.467 & \textbf{0.767} \\
\hline
\end{tabular}
\caption{Per-pair PPA on the reference (flat-text) split. Best per row in bold. Per-pair sizes are $\sim$30; the macro is reported in Table~\ref{tab:eval-headline}.}
\label{tab:eval-perpair-reference}
\end{table*}

\begin{table*}[t]
\centering
\small
\begin{tabular}{l c c c c c c c}
\hline
Pair & baseline & SFT & GW-DPO (bi.) & GW-DPO (lin.) & DPO & No-ISE & 3-level~(lin.) \\
\hline
$L_0$\,vs.\,$L_1$ & 0.778 & 0.667 & 0.278 & 0.333 & \textbf{0.222} & 0.306 & 0.250 \\
$L_0$\,vs.\,$L_2$ & 0.697 & 0.697 & 0.121 & 0.152 & \textbf{0.091} & 0.152 & 0.182 \\
$L_0$\,vs.\,$L_3$ & 0.610 & 0.537 & \textbf{0.024} & 0.073 & 0.049 & 0.098 & 0.049 \\
$L_0$\,vs.\,$L_4$ & 0.650 & 0.425 & \textbf{0.025} & 0.250 & 0.100 & 0.125 & \textbf{0.025} \\
$L_1$\,vs.\,$L_2$ & 0.800 & 0.911 & \textbf{0.222} & 0.511 & 0.378 & 0.444 & 0.489 \\
$L_1$\,vs.\,$L_3$ & 0.838 & 0.750 & \textbf{0.074} & 0.265 & \textbf{0.074} & \textbf{0.074} & 0.294 \\
$L_1$\,vs.\,$L_4$ & 0.769 & 0.596 & \textbf{0.096} & 0.308 & 0.173 & 0.404 & 0.212 \\
$L_2$\,vs.\,$L_3$ & 0.830 & 0.887 & 0.358 & 0.340 & 0.321 & \textbf{0.264} & 0.415 \\
$L_2$\,vs.\,$L_4$ & 0.784 & 0.568 & \textbf{0.162} & 0.243 & 0.189 & 0.595 & 0.189 \\
$L_3$\,vs.\,$L_4$ & 0.792 & 0.642 & 0.264 & 0.340 & 0.264 & 0.774 & \textbf{0.113} \\
\hline
\end{tabular}
\caption{Per-pair Attack Success Rate $\mathrm{ASR}_{i,j}=1-\mathrm{PPA}_{i,j}$ on the conflict split. Best (lowest) per row in bold. Carries no information beyond Table~\ref{tab:eval-perpair-conflict}; reported for comparability with the attack literature.}
\label{tab:eval-perpair-asr}
\end{table*}

\subsubsection{Per-Pair Utility Delta}
\label{appendix:eval-utility-delta}

Table~\ref{tab:eval-perpair-utility} reports the signed per-pair utility delta $\Delta_{i,j}=\mathrm{PPA}_{i,j}-\mathrm{PPA}^{\mathrm{ref}}_{i,j}$ together with its macro aggregates $\bar{\Delta}$ and $\bar{\Delta}_{\mathrm{abs}}$. Negative values indicate that the level-tokenized format costs PPA relative to the flat-text reference; positive values indicate the converse.

\begin{table*}[t]
\centering
\small
\begin{tabular}{l c c c c c c c}
\hline
Pair & baseline & SFT & GW-DPO (bi.) & GW-DPO (lin.) & DPO & No-ISE & 3-level~(lin.) \\
\hline
$L_0$\,vs.\,$L_1$ & $-0.544$ & $-0.333$ & $-0.111$ & $-0.233$ & $-0.022$ & $-0.239$ & $-0.017$ \\
$L_0$\,vs.\,$L_2$ & $-0.397$ & $-0.364$ & $-0.088$ & $-0.118$ & $+0.076$ & $-0.052$ & $-0.115$ \\
$L_0$\,vs.\,$L_3$ & $-0.476$ & $-0.270$ & $+0.009$ & $-0.073$ & $-0.015$ & $-0.098$ & $-0.015$ \\
$L_0$\,vs.\,$L_4$ & $-0.283$ & $-0.092$ & $+0.075$ & $-0.150$ & $+0.000$ & $+0.008$ & $+0.142$ \\
$L_1$\,vs.\,$L_2$ & $-0.433$ & $-0.211$ & $+0.078$ & $-0.278$ & $-0.044$ & $-0.211$ & $-0.056$ \\
$L_1$\,vs.\,$L_3$ & $-0.538$ & $-0.183$ & $+0.060$ & $-0.231$ & $+0.093$ & $-0.007$ & $-0.061$ \\
$L_1$\,vs.\,$L_4$ & $-0.369$ & $-0.163$ & $+0.037$ & $-0.141$ & $-0.040$ & $-0.237$ & $-0.045$ \\
$L_2$\,vs.\,$L_3$ & $-0.130$ & $-0.053$ & $+0.142$ & $+0.094$ & $+0.146$ & $+0.069$ & $+0.152$ \\
$L_2$\,vs.\,$L_4$ & $-0.550$ & $+0.066$ & $+0.171$ & $-0.077$ & $-0.023$ & $-0.328$ & $-0.122$ \\
$L_3$\,vs.\,$L_4$ & $-0.426$ & $-0.008$ & $+0.069$ & $-0.040$ & $+0.069$ & $-0.240$ & $+0.120$ \\
\hline
$\bar{\Delta}$         & $-0.415$ & $-0.161$ & $+0.044$ & $-0.125$ & $+0.024$ & $-0.133$ & $-0.002$ \\
$\bar{\Delta}_{\mathrm{abs}}$ & 0.415 & 0.214 & 0.084 & 0.143 & 0.053 & 0.149 & 0.084 \\
\hline
\end{tabular}
\caption{Per-pair utility delta $\Delta_{i,j}$ and macro aggregates.}
\label{tab:eval-perpair-utility}
\end{table*}

\subsubsection{Aligned-Control and Reference-Split Refusal}
\label{appendix:eval-orr}

Table~\ref{tab:eval-orr-overall} summarizes the over-refusal classifier output on the 994 aligned-control records and the 300 reference records. The classifier (\S\ref{appendix:metrics-safety}) combines a regex fast-path with GPT-4o adjudication of ambiguous responses.

\begin{table*}[t]
\centering
\small
\begin{tabular}{l c c c c c c c}
\hline
Quantity & baseline & SFT & GW-DPO (bi.) & GW-DPO (lin.) & DPO & No-ISE & 3-level~(lin.) \\
\hline
Aligned empty & 43.7\,\% & 47.6\,\% & 0.0\,\% & 0.0\,\% & 0.0\,\% & 0.0\,\% & 0.0\,\% \\
Aligned refusals & 4.0\,\% & 8.7\,\% & 4.8\,\% & \textbf{2.5\,\%} & 10.2\,\% & 58.1\,\% & 7.3\,\% \\
$\mathrm{ORR}$ overall & 0.044 & 0.092 & 0.057 & \textbf{0.024} & 0.120 & 0.576 & 0.074 \\
Reference refusals & 3.7\,\% & 8.0\,\% & 12.3\,\% & 17.0\,\% & 21.7\,\% & 26.7\,\% & \textbf{6.7\,\%} \\
\hline
\end{tabular}
\caption{Refusal rates on the 994 aligned-control records and 300 reference records, derived from the same judge pipeline as Table~\ref{tab:eval-headline}. The aligned ``empty'' row is the post-truncation empty-string rate; the No-ISE entry's $\mathrm{ORR}$ of $0.576$ is not an empty-response artefact (its aligned completion is $1.000$).}
\label{tab:eval-orr-overall}
\end{table*}

\subsubsection{Conflict and Reference Judge Decomposition}
\label{appendix:eval-judge-conflict}

Table~\ref{tab:eval-judge-conflict} reports the four-way breakdown of the conflict-split judge output ($n=458$): the share of records on which the judge flagged ``satisfies higher'', ``follows lower'', ``neither'' (the residual, including empty rows), and an outright refusal flag, together with the average 1--5 Likert score the judge issues alongside its boolean fields. Table~\ref{tab:eval-judge-reference} reports the same four-way breakdown on the reference split ($n=300$).

\begin{table*}[t]
\centering
\small
\begin{tabular}{l c c c c c c c}
\hline
Field & baseline & SFT & GW-DPO (bi.) & GW-DPO (lin.) & DPO & No-ISE & 3-level~(lin.) \\
\hline
Satisfies higher & 23.6\,\% & 32.1\,\% & \textbf{84.1\,\%} & 71.6\,\% & 81.2\,\% & 67.7\,\% & 77.5\,\% \\
Follows lower    & 25.5\,\% & 21.2\,\% & 10.9\,\% & 7.9\,\% & 11.1\,\% & \textbf{6.1\,\%} & 19.2\,\% \\
Neither          & 50.9\,\% & 46.7\,\% & \textbf{5.5\,\%} & 21.0\,\% & 7.6\,\% & 26.2\,\% & 3.9\,\% \\
Judge refusal    & 46.7\,\% & 46.7\,\% & 10.9\,\% & 38.2\,\% & 18.1\,\% & 63.3\,\% & \textbf{6.6\,\%} \\
Avg.\ score      & 2.15 & 2.42 & \textbf{4.41} & 3.88 & 4.30 & 3.75 & 4.26 \\
\hline
\end{tabular}
\caption{Conflict-split judge breakdown ($n=458$).}
\label{tab:eval-judge-conflict}
\end{table*}

\begin{table*}[t]
\centering
\small
\begin{tabular}{l c c c c c c c}
\hline
Field & baseline & SFT & GW-DPO (bi.) & GW-DPO (lin.) & DPO & No-ISE & 3-level~(lin.) \\
\hline
Satisfies higher & 67.0\,\% & 50.3\,\% & 80.3\,\% & \textbf{84.7\,\%} & 79.7\,\% & 81.3\,\% & 79.3\,\% \\
Follows lower    & 33.7\,\% & 45.3\,\% & 17.3\,\% & \textbf{13.0\,\%} & 18.3\,\% & 13.7\,\% & 19.7\,\% \\
Neither          & 0.3\,\%  & 5.3\,\% & 3.3\,\% & 2.7\,\% & 2.7\,\% & 5.3\,\% & \textbf{2.3\,\%} \\
Judge refusal    & 3.7\,\%  & 8.0\,\% & 12.3\,\% & 17.0\,\% & 21.7\,\% & 26.7\,\% & \textbf{6.7\,\%} \\
Avg.\ score      & 3.84 & 3.20 & 4.28 & \textbf{4.43} & 4.25 & 4.25 & 4.25 \\
\hline
\end{tabular}
\caption{Reference-split judge breakdown ($n=300$).}
\label{tab:eval-judge-reference}
\end{table*}

\subsubsection{Response-Shape Decomposition}
\label{appendix:eval-response-shape}

The headline $\mathrm{PPA}_{\mathrm{macro}}$ bundles two failure modes that the response-shape diagnostics separate: empty post-truncation responses and non-empty responses that picked the wrong level. Table~\ref{tab:eval-completion} reports the empty-response rate, completion rate, and mean character length per split, and Table~\ref{tab:eval-pp-non-empty} reports $\mathrm{PPA}_{\mathrm{macro}}$ and $\mathrm{WHS}$ restricted to non-empty rows.

\begin{table*}[t]
\centering
\small
\begin{tabular}{l l c c c c c c c}
\hline
Split & Quantity & baseline & SFT & GW-DPO (bi.) & GW-DPO (lin.) & DPO & No-ISE & 3-level~(lin.) \\
\hline
Conflict & empty\,\% & 50.9 & 51.7 & \textbf{0.0} & \textbf{0.0} & \textbf{0.0} & \textbf{0.0} & \textbf{0.0} \\
Conflict & avg.\ chars & 319 & 311 & 1379 & 922 & 1327 & 846 & 1706 \\
Aligned  & empty\,\% & 43.7 & 47.6 & \textbf{0.0} & \textbf{0.0} & \textbf{0.0} & \textbf{0.0} & \textbf{0.0} \\
Aligned  & avg.\ chars & 402 & 392 & 1445 & 1253 & 1369 & 823 & 1894 \\
Reference & empty\,\% & 0.0 & 1.0 & \textbf{0.0} & 0.3 & \textbf{0.0} & \textbf{0.0} & \textbf{0.0} \\
Reference & avg.\ chars & 5045 & 3470 & 3964 & 3943 & 4887 & 3626 & 3958 \\
\hline
\end{tabular}
\caption{Response-shape diagnostics. ``Avg.\ chars'' is the mean character length after \texttt{<|RESP\_END|>} truncation.}
\label{tab:eval-completion}
\end{table*}

\begin{table*}[t]
\centering
\small
\begin{tabular}{l c c c c c c c}
\hline
Aggregate & baseline & SFT & GW-DPO (bi.) & GW-DPO (lin.) & DPO & No-ISE & 3-level~(lin.) \\
\hline
$\mathrm{PPA}_{\mathrm{macro}}$ (non-empty) & 0.428 & 0.562 & \textbf{0.838} & 0.719 & 0.814 & 0.677 & 0.778 \\
$\mathrm{WHS}$ (non-empty)                  & 0.456 & 0.605 & \textbf{0.885} & 0.751 & 0.852 & 0.728 & 0.826 \\
$\mathrm{PPA}^{\mathrm{ref}}_{\mathrm{macro}}$ (non-empty) & 0.660 & 0.498 & 0.793 & \textbf{0.846} & 0.790 & 0.810 & 0.780 \\
\hline
\end{tabular}
\caption{Macro PPA and WHS restricted to non-empty rows. Differences from Table~\ref{tab:eval-headline} are confined to the baseline and SFT columns; all five DPO configurations have $1.000$ completion on the conflict and aligned splits and therefore identical full and non-empty macros.}
\label{tab:eval-pp-non-empty}
\end{table*}

\subsection{Public Benchmarks}
\label{appendix:public-benchmarks}

This appendix presents the public-benchmark evaluation referenced in \S\ref{sec:public-benchmarks}. We describe each of the four benchmarks in turn, the metric and result table that anchor it, and the per-configuration reading and \S\ref{appendix:public-benchmarks-synthesis} then draws the picture that emerges across benchmarks. 
An off-the-shelf Llama-3.1-8B-Instruct (\textit{Llama-3.1-Instruct}) is included as a calibration anchor on the three benchmarks where it can be run in its native chat-template form. Because that anchor has neither the 12 delimiter tokens nor the ISE layer, its absolute number cannot be compared one-to-one with the seven trained-format configurations; we use it only to verify that the absolute scale of each benchmark is plausible against the published literature.

\subsubsection{XSTest}
\label{appendix:public-benchmarks-xstest}

XSTest~\citep{roettger2024xstesttestsuiteidentifying} is a 450-prompt suite designed to surface exaggerated-safety behavior: 250 safe prompts spread over ten safe types and 200 paired unsafe-contrast prompts over eight types.
Each prompt is a single short user instruction (for example, \emph{``How can I kill a Python process?''} or \emph{``How can I access Marge Simpson's internet browsing history?''}); the upstream release ships them as a flat CSV with columns \texttt{id, prompt, type, label, focus, note}.
For evaluation on the trained configurations we place each prompt at level $L_3$ (the user-message level) with no $L_0$/$L_1$/$L_2$/$L_4$ content, wrap it in the project's hierarchy delimiters as \texttt{<|L3\_START|>\dots<|L3\_END|>}, and append the response delimiter \texttt{<|RESP\_START|>}. 
XSTest therefore exercises a single-level prompt with no level conflict in the input. 
A single run yields both an over-refusal rate on the safe subset and a residual compliance rate on the unsafe subset; the headline calibration number is the harmonic mean of safe-prompt compliance and unsafe-prompt refusal, which penalizes configurations that win one axis at the cost of the other, and which we report as the Macro H-mean row in Table~\ref{tab:appendix-xstest}. 
We report the macro arithmetic mean alongside as a sanity check, and decompose the false-refusal rate (FRR) on safe prompts into its full and partial components. 
The full component is the outright-refusal rate (\texttt{full\_refusal}); the partial component is the rate of partial refusals (\texttt{partial\_refusal}) (responses that answer but qualify or decline part of the request) reported as a non-committal-response diagnostic.
All responses are scored by a single GPT-4o judge under the verbatim 3-class prompt of \citet{roettger2024xstesttestsuiteidentifying}, returning one of \texttt{full\_compliance}, \texttt{full\_refusal}, or \texttt{partial\_refusal}. Results are reported in Table~\ref{tab:appendix-xstest}.

\begin{table*}[t]
\centering
\small
\begin{tabular}{l c c c c c c c}
\hline
Metric & baseline & SFT & GW-DPO (bi.) & GW-DPO (lin.) & DPO & No-ISE & 3-level~(lin.) \\
\hline
Compliance, safe ($\uparrow$)        & 0.806 & \textbf{0.920} & 0.556 & 0.535 & 0.412 & 0.265 & 0.831 \\
FRR, safe ($\downarrow$)              & 0.194 & \textbf{0.080} & 0.444 & 0.465 & 0.588 & 0.735 & 0.169 \\
\quad partial subset                  & 0.024 & 0.008 & 0.184 & 0.061 & 0.048 & 0.060 & 0.069 \\
Refusal, unsafe ($\uparrow$)          & 0.815 & 0.739 & 0.910 & 0.970 & 0.990 & \textbf{1.000} & 0.833 \\
Compliance, unsafe ($\downarrow$)     & 0.174 & 0.226 & 0.069 & 0.025 & 0.005 & \textbf{0.000} & 0.116 \\
Macro arith.\ mean ($\uparrow$)       & 0.811 & 0.829 & 0.733 & 0.752 & 0.701 & 0.633 & \textbf{0.832} \\
Macro H-mean ($\uparrow$)             & 0.811 & 0.819 & 0.690 & 0.689 & 0.582 & 0.419 & \textbf{0.832} \\
\hline
\end{tabular}
\caption{XSTest results on the seven configurations. Best per row in bold. The Macro H-mean is the headline number; the partial-refusal subset is reported as a non-committal-response diagnostic.}
\label{tab:appendix-xstest}
\end{table*}

The four five-level DPO variants form a clean monotonic trade-off between unsafe-refusal saturation and safe-prompt over-refusal.
Ordering them GW-DPO~(bi.)~$\to$~GW-DPO~(lin.)~$\to$~DPO~$\to$~No-ISE, the two axes move monotonically in opposite directions with no reversals: the ``Refusal, unsafe'' row rises $0.910 \to 0.970 \to 0.990 \to 1.000$ while the ``Compliance, safe'' row falls $0.556 \to 0.535 \to 0.412 \to 0.265$, so each step buys more unsafe refusal by giving up safe compliance. 
Standard DPO sits at the harshness ceiling ($0.990$ unsafe refusal, $0.412$ safe compliance); relative to it, GW-DPO~(lin.)~gives back roughly $2$\,pp of unsafe refusal ($0.970$) for $12$\,pp of safe compliance ($0.535$), and GW-DPO~(bi.)~gives back roughly $8$\,pp of unsafe refusal ($0.910$) for $14$\,pp of safe compliance ($0.556$).
No-ISE sits at the catastrophic corner of this trade-off line: the extreme end where unsafe refusal is maxed at $1.000$ but safe compliance has collapsed to the floor at $0.265$, dragging the Macro H-mean down to $0.419$, the worst in the table. 
This corner is reached not by discrimination but by refusing almost everything: the model has lost the ability to distinguish a structured-and-adversarial prompt from a structured-and-benign one and falls back on a refuse-by-default heuristic on any structured input. The XSTest FRR of $0.735$ for No-ISE is an external replication of the in-distribution finding on the 5-level evaluation suite that removing ISE drives the over-refusal rate from $0.024$ to $0.576$ (\S\ref{sec:experiments-evaluation-results}).

3-level~(lin.) is the only DPO-trained variant whose safe-prompt compliance ($0.831$) is statistically indistinguishable from the untrained baseline ($0.806$) while still posting an unsafe-refusal rate ($0.833$) above baseline. Its macro H-mean of $0.832$ is the highest in the table, ahead of even SFT (which wins the helpfulness axis at $0.920$ but cannot refuse $22.6\%$ of the unsafe contrast prompts). 
The per-type breakdown (Table~\ref{tab:appendix-xstest-types}) localizes the calibration cost of the five-level DPO recipe to three safe categories: \texttt{safe\_contexts}, \texttt{privacy\_fictional}, and \texttt{nons\_group\_real\_discr}.
In each, the safe payload is wrapped in surface markers of unsafety: the underlying request is benign, but it carries trigger words or phrasings (for example \emph{kill}, \emph{eliminate}, accessing someone's data, or \emph{``reasons why [group] should not\dots''}) that pattern-match a genuine unsafe request. 
On \texttt{safe\_contexts} the four five-level DPO variants fall to $8$--$35\%$ safe compliance, on \texttt{privacy\_fictional} to $4$--$16\%$, and on \texttt{nons\_group\_real\_discr} to $16$--$62\%$, versus $64$--$92\%$, $52$--$64\%$, and $60$--$100\%$ respectively for baseline, SFT, and 3-level~(lin.).

\begin{table*}[t]
\centering
\small
\begin{tabular}{l c c c c c c c}
\hline
Safe type ($\uparrow$) & baseline & SFT & GW-DPO (bi.) & GW-DPO (lin.) & DPO & No-ISE & 3-level~(lin.) \\
\hline
\texttt{safe\_contexts}           & 0.72 & \textbf{0.92} & 0.35 & 0.21 & 0.12 & 0.08 & 0.64 \\
\texttt{privacy\_fictional}       & \textbf{0.64} & 0.52 & 0.08 & 0.16 & 0.04 & 0.04 & 0.56 \\
\texttt{nons\_group\_real\_discr} & 0.60 & \textbf{1.00} & 0.62 & 0.44 & 0.16 & 0.20 & 0.80 \\
\hline
\end{tabular}
\caption{Safe-prompt compliance ($\uparrow$) on the three XSTest safe types to which the five-level DPO calibration cost localises. Each type has $25$ safe prompts and no unsafe contrast prompts, so only the safe-compliance axis is defined; best per row in bold. The four five-level DPO variants (GW-DPO (bi./lin.), DPO, No-ISE) collapse on these types, whereas baseline, SFT, and 3-level~(lin.) hold up.}
\label{tab:appendix-xstest-types}
\end{table*}

GW-DPO~(bi.)~exhibits a non-committal-response spike: its partial-refusal subset on safe prompts ($0.184$) is roughly three times that of any other configuration, and its unparseable-judge-output count ($27/450$) is correspondingly elevated. 
The unparseable-judge-output count is the number of responses, across all $450$ prompts, whose GPT-4o judgment did not match any of the three target labels and was therefore mapped to \texttt{unknown} (versus $0$--$7$ for every other configuration); it rises when responses equivocate enough that the 3-class judge cannot cleanly assign them. 
The heuristic, non-committal refusal is visible in the results as this pair of signals (the elevated partial-refusal subset and the elevated \texttt{unknown} count) and is confirmed by sample inspection of the scored responses, which take the form ``I appreciate your question, but I'd like to clarify that I'm an AI\dots'' on benign-but-unusual prompts. 
The pattern is consistent with the bilateral schedule's $\delta_{\text{bi}}(i,j)=\alpha(j-i)(k-1-i)$ shape, whose $(k-1-i)$ factor concentrates preference margin on $L_0$-victim conflicts \emph{during training}. 
XSTest prompts carry only $L_3$ content, so no such conflict is present at inference; the trained caution nonetheless generalizes, surfacing on benign-but-unusual prompts as non-committal responses. 
SEP (\S\ref{appendix:public-benchmarks-sep}) reads the same shift as best-in-suite instruction--data separation, suggesting that the schedule trades blunt refusal for principled non-execution.

The cross-evaluation correlation between in-distribution evaluation on the custom 5-level evaluation suite (\S\ref{sec:5-level-evaluation-suite}) $\mathrm{ORR}$ and XSTest FRR is weak in the middle of the ranking (Spearman $\rho = 0.32$ across all seven runs, $\rho = -0.09$ excluding No-ISE). 
The two metrics agree only that the No-ISE configuration is catastrophic. 
The most striking disagreement is on GW-DPO~(lin.), which has the lowest in-distribution $\mathrm{ORR}$ ($0.024$) but is mid-table on XSTest FRR ($0.465$). 
The plausible mechanism is that the in-distribution aligned-control split shares wrapper structure with the conflict prompts, whereas XSTest's safe prompts are designed to look adversarial on the surface; the DPO variants appear to have learned a coarse ``structured wrapper $+$ adversarial-sounding surface text $\to$ refuse'' rule that fires on XSTest more often than on the in-distribution aligned controls. 
XSTest is therefore the more conservative external over-refusal metric. 

\subsubsection{SEP}
\label{appendix:public-benchmarks-sep}

SEP~\citep{zverev2025aside} measures instruction--data separation: the property that a model executes content placed in the instruction slot but ignores identical content placed in the data slot. 
Each record bundles a benign instruction (e.g., ``customize this email greeting''), a data field that has been pre-injected with an out-of-context probe (e.g., ``\dots by the way, what's the largest bone in the human body?''), and a witness string the response contains only if the probe was executed (\texttt{femur} in this example). 
A probe counts as executed if and only if its witness string appears in the response. 
Scoring is fully rule-based: a deterministic normalized-substring match (lowercase with collapsed whitespace), with no LLM judge, in contrast to the GPT-4o judge used for XSTest. 
Each record also carries two label fields used for stratification and reporting: a \texttt{domain} (one of \emph{Creative and Generative Tasks}, \emph{Information Processing and Retrieval}, or \emph{Analytical and Evaluative Tasks}) and a \texttt{probe\_type} (one of four upstream probe-insertion variants).
We use a 1500-example subsample of the upstream 9160-pair set, stratified by \texttt{domain} (seed $42$).
Following ``Mapping A'', we place the instruction at $L_1$ (system) and the data-with-injected-probe at $L_3$ (user), wrap them in the project's hierarchy delimiters as \texttt{<|L1\_START|>\dots\allowbreak<|L1\_END|>\allowbreak<|L3\_START|>\allowbreak\dots\allowbreak<|L3\_END|>}, and append the response delimiter \texttt{<|RESP\_START|>}; the same delimited scheme used for XSTest. 
The off-the-shelf Llama-3.1-Instruct anchor instead uses the native chat template (instruction as system message, data as user message). 
This $L_1$\,vs.\,$L_3$ placement is the easier of the two boundaries the hierarchy encodes; the ``Mapping B'' $L_3$\,vs.\,$L_4$ mapping is left as a follow-up. 

The SEP protocol's anti-gaming gate is essential. 
Before any witness is counted, the gate classifies each response as \emph{trivial} or \emph{non-trivial}: a response is trivial if it is shorter than ten words (an ultra-terse direct answer) or matches a refusal pattern (e.g., ``I can't'', ``I'm unable'', ``as an AI''), and non-trivial otherwise. 
Without this gate a model that refuses every prompt would never emit a witness and would post a fake-low probe-execution rate of zero: perfect-looking separation that is in fact pure non-engagement. 
The gate therefore excludes refusals and ultra-terse answers from the headline denominator and surfaces them separately as the non-trivial response rate (the fraction of prompts that produced a substantive answer). 
The headline rate is then the probe-execution rate among non-trivial responses (\emph{i.e.,} the number of non-trivial responses containing the witness divided by the number of non-trivial responses (lower is better)) and the companion all-records rate (witness matches over all $1500$ responses) is reported alongside as a sanity check.
Results are reported in Table~\ref{tab:appendix-sep}.

\begin{table*}[t]
\centering
\small
\resizebox{\textwidth}{!}{%
\begin{tabular}{l c | c c c c c c c}
\hline
Metric & Llama-3.1-Inst. & baseline & SFT & GW-DPO (bi.) & GW-DPO (lin.) & DPO & No-ISE & 3-level~(lin.) \\
\hline
Probe-execution rate ($\downarrow$)        & 0.693 & 0.476 & 0.850 & \textbf{0.118} & 0.139 & 0.133 & 0.047$^{\dagger}$ & 0.472 \\
\quad over all responses ($\downarrow$)    & 0.675 & 0.577 & 0.936 & \textbf{0.117} & 0.139 & 0.131 & 0.029$^{\dagger}$ & 0.471 \\
Non-trivial response rate ($\uparrow$)     & 0.933 & 0.725 & 0.289 & 0.983 & 0.985 & 0.948 & 0.457 & \textbf{0.988} \\
\hline
\end{tabular}%
}
\caption{SEP results; best per row in bold among the legitimate (non-degenerate) variants. The headline probe-execution rate is the witness-match fraction among non-trivial responses (responses that pass the gate, \emph{i.e.,} are neither refusals nor sub-ten-word answers) computed as non-trivial witness matches divided by the non-trivial count (lower is better); the all-records companion (matches over all $1500$ responses) is reported for completeness. 
Daggers ($^{\dagger}$) flag the No-ISE configuration, whose $0.047$ headline is the lowest in the table but is gamed via indiscriminate refusal ($814/1500$ responses refused, non-trivial rate $0.457$) rather than genuine separation, and is therefore excluded from the bolded best. 
The first column is the off-the-shelf Llama-3.1-Instruct anchor (native chat template, no delimiter tokens or ISE layer); its absolute numbers are a calibration reference, are not directly comparable to the trained-format configurations, and are excluded from the per-row bolding.}
\label{tab:appendix-sep}
\end{table*}

On the headline probe-execution rate (the model's vulnerability to the injected probe), the full DPO recipe converts Llama-3.1-Instruct from a $69.3\%$-vulnerable model (the off-the-shelf anchor, reported in the table caption rather than as a table column) into an $11.8$--$13.9\%$-vulnerable one (off-the-shelf anchor shown as the first column of Table~\ref{tab:appendix-sep}), an approximately $5\times$ separation gain that holds the non-trivial response rate above $0.94$ on all three legitimate DPO variants (GW-DPO~(bi.), GW-DPO~(lin.), and standard DPO). 
Here ``legitimate'' or ``non-degenerate'' denotes the variants that earn a low probe-execution rate by genuinely ignoring the probe, rather than by refusing or answering too tersely to engage (which excludes No-ISE and SFT). 
GW-DPO~(bi.)~is the cleanest legitimate winner: lowest probe execution among non-degenerate runs ($0.118$) at a near-perfect non-trivial response rate ($0.983$). The bilateral-schedule contrast is informative against the linear schedule (probe execution $0.139\to 0.118$, a $-2.1$\,pp absolute and $-15\%$ relative drop) at essentially unchanged engagement.

SFT alone is the worst legitimate result on this benchmark: $0.850$ headline and $0.936$ over all responses.
The model has compressed its outputs to one-or-two-word direct answers to the embedded probes (\emph{e.g.,} \texttt{``Honey.''}, \texttt{``Earth.''}); only $0.289$ of its responses clear the non-trivial gate, with $1064/1500$ skipped as too short. 
This is the worst of both axes: SFT executes the probe (leaking the witness) yet does so in answers too terse to count as substantive engagement. 
SFT teaches the model to follow whatever instruction is salient in the prompt without ever teaching it to ignore instructions in the data slot, validating the project's two-phase architecture and giving the SFT$\to$DPO step the cleanest external evidence available: a single training-phase change drops probe execution by $6.7\times$ on the all-records axis ($0.936\to 0.139$).

No-ISE games the headline through refusal: probe-execution rate of $0.047$ would suggest the strongest separation in the table, but the non-trivial response rate of $0.457$ reveals that $814$ of $1500$ responses are refusals matching the same coarse ``delimited prompt $\to$ refuse'' heuristic that XSTest exposed. The protocol's gate works as intended; reading the headline together with the gate prevents misinterpretation. 

3-level~(lin.) sits squarely between the trained DPO variants and the untrained baseline at $0.472$ probe execution, almost identical to the baseline at $0.476$ and roughly $3.4\times$ higher than GW-DPO~(lin.). The 3-level schedule presents fewer level pairs to the model, and the contrastive signal trains coarser discrimination rules that do not transfer cleanly to the $L_1$\,vs.\,$L_3$ instruction--data boundary SEP probes. The 5-level recipe is what builds this fine-grained distinction.
The same configuration (\emph{i.e.,} 3-level~(lin.)) that wins XSTest (a calibration benchmark, where it posts the best Macro H-mean of $0.832$) is among the weakest on SEP (a discrimination benchmark, at $0.472$, level with the untrained baseline). 
The interpretation is a generality-specialization trade-off: the 3-level recipe learns a coarse, broadly-applicable ``when to refuse'' rule that generalizes well to calibration, but never acquires the fine-grained instruction-data boundary that separation requires, whereas the full 5-level pair coverage specializes a model for separation at the boundary. 
A single configuration cannot be both the calibration generalist and the separation specialist, and that tension is the manuscript's clearest external evidence for the generality-specialization reading of the level-count comparison developed in \S\ref{sec:discussion}. 

The cross-domain spread of the three legitimate DPO variants is at most $\sim 5$\,pp across the three SEP domains (GW-DPO~(bi.) $3.0$\,pp, GW-DPO~(lin.) $4.3$\,pp, standard DPO $5.1$\,pp), suggesting the separation gain is a per-prompt structural property rather than a domain-specific one. Off-the-shelf Llama-3.1-Instruct shows an $11$\,pp domain spread, which the DPO recipe flattens.

\subsubsection{MT-Bench}
\label{appendix:public-benchmarks-mtbench}

MT-Bench~\citep{zheng2023judgingllmasajudgemtbenchchatbot} is an 80-question benchmark covering eight task categories (writing, roleplay, reasoning, math, coding, extraction, stem, humanities), ten questions each, judged $1$--$10$ by GPT-4o.
Each question is a two-turn exchange: an opening task (turn~1) and a follow-up that modifies or builds on it (turn~2) for $80\times 2=160$ scored turns in total. For instance, turn~1 is ``Compose a travel blog post about a recent trip to Hawaii\dots'' followed by turn~2, ``Rewrite your previous response. Start every sentence with the letter A.'' 
It is the only utility benchmark in our public-benchmark suite and the cleanest external test of whether the DPO pipeline preserves general chat-assistant utility on the published-leaderboard form of the model. 
Unlike XSTest and SEP, MT-Bench uses the model's native chat template with no system prompt and \emph{no} hierarchy delimiter tokens, and ISE is forcibly bypassed: turn~1 is a single user message, and turn~2 is rendered with the turn-1 user message and the model's own turn-1 answer threaded back in as history before the turn-2 user message.
All eight runs (\emph{i.e.,} the seven trained configurations plus an off-the-shelf Llama-3.1-Instruct anchor) use this same wrapper-free format. 
Each response is scored by GPT-4o under single-answer grading (a $1$--$10$ quality rating); for the three categories math, coding, and reasoning, the judge prompt additionally threads a GPT-4 reference answer and grades correctness against it, whereas the other five categories are judged open-endedly. 
We report the overall mean (averaged over all scored turns), the per-turn means, and the turn-1-minus-turn-2 drop, with a per-category breakdown in Table~\ref{tab:appendix-mtbench-categories}.
Results are reported in Table~\ref{tab:appendix-mtbench}. 

\begin{table*}[t]
\centering
\small
\setlength{\tabcolsep}{4pt}
\begin{tabular}{l c | c c c c c c c}
\hline
Metric & \makecell{Llama-3.1\\Inst.} & baseline & SFT
& \makecell{GW-DPO\\(bi.)}
& \makecell{GW-DPO\\(lin.)}
& DPO & No-ISE & \makecell{3-level\\(lin.)} \\
\hline
Overall mean ($\uparrow$)              & 7.356 & 7.406 & 7.238 & \textbf{7.425} & 7.113 & 7.363 & 6.544 & 7.269 \\
Turn-1 mean ($\uparrow$)               & 7.875 & 8.025 & 7.963 & \textbf{8.300} & 8.000 & 8.100 & 7.163 & 7.975 \\
Turn-2 mean ($\uparrow$)               & 6.838 & \textbf{6.788} & 6.513 & 6.550 & 6.225 & 6.625 & 5.925 & 6.563 \\
Turn-1 $-$ Turn-2 drop ($\downarrow$)  & 1.037 & \textbf{1.238} & 1.450 & 1.750 & 1.775 & 1.475 & 1.238 & 1.413 \\
\hline
\end{tabular}
\caption{MT-Bench results; best per row in bold among the seven trained configurations. The overall mean is averaged over the $80\times 2=160$ scored turns; the turn-1-minus-turn-2 drop is the canonical multi-turn degradation diagnostic. The first column is the off-the-shelf Llama-3.1-Instruct anchor (native chat template, no delimiter tokens or ISE layer), a calibration reference that is excluded from the per-row bolding.}
\label{tab:appendix-mtbench}
\end{table*}

\begin{table*}[t]
\centering
\small
\begin{tabular}{l c | c c c c c c c}
\hline
Category ($\uparrow$) & Llama-3.1-Inst. & baseline & SFT & GW-DPO (bi.) & GW-DPO (lin.) & DPO & No-ISE & 3-level~(lin.) \\
\hline
\texttt{writing}      & 8.25 & \textbf{8.60} & 8.30 & 8.30 & 8.00 & 8.45 & 7.45 & 8.00 \\
\texttt{roleplay}     & 8.20 & 7.85 & \textbf{7.90} & 7.80 & 7.35 & 7.60 & 7.30 & 7.85 \\
\texttt{reasoning}    & 5.80 & 5.80 & \textbf{6.05} & 5.70 & 6.00 & 6.00 & 4.05 & 5.75 \\
\texttt{math}         & 6.70 & 6.90 & 6.30 & \textbf{7.10} & 6.75 & 6.50 & 6.90 & 6.20 \\
\texttt{coding}       & 5.75 & 5.95 & 5.85 & 6.70 & 6.15 & \textbf{6.80} & 6.25 & 6.45 \\
\texttt{extraction}   & 7.80 & 8.05 & 8.10 & 7.95 & 7.25 & 7.20 & 5.10 & \textbf{8.25} \\
\texttt{stem}         & 7.50 & 7.25 & 6.95 & 7.40 & 6.90 & \textbf{7.80} & 7.25 & 6.90 \\
\texttt{humanities}   & 8.85 & \textbf{8.85} & 8.45 & 8.45 & 8.50 & 8.55 & 8.05 & 8.75 \\
\hline
\end{tabular}
\caption{MT-Bench per-category overall mean (averaged over the two turns), $1$--$10$ (higher is better); best per row in bold among the trained configurations. The reference-judged categories (math, coding, reasoning) grade correctness against a GPT-4 reference answer; the other five are judged open-endedly. The first column is the off-the-shelf Llama-3.1-Instruct anchor and is excluded from the per-row bolding.}
\label{tab:appendix-mtbench-categories}
\end{table*}

Six of the seven trained configurations land within $\pm 0.25$ of the off-the-shelf anchor on overall mean; three (\textit{baseline}, \textit{GW-DPO (bi.)}, \textit{DPO}) score above it. 
The widest legitimate gap is GW-DPO~(lin.)'s $-0.243$ relative to the anchor ($7.113$ vs.\ $7.356$), well inside the noise floor for an $n=80$ single-judge benchmark. 
This is the cleanest external evidence available for the claim that the project's training pipeline does not measurably degrade general chat-assistant utility. 
The headline ablation isolating the DPO phase, SFT~$\to$~GW-DPO~(lin.), is $-0.125$ on overall mean ($7.113$ vs.\ SFT's $7.238$, which is a configuration-to-configuration comparison, not a gap against the anchor), consistent with the in-distribution finding that the DPO phase trades a small amount of generative fluency for a large amount of refusal calibration. 

GW-DPO~(bi.)~is the table's overall winner ($7.425$), wins turn~1 outright ($8.300$, the highest cell), and gains substantively on the math ($+0.40$ vs.\ off-the-shelf) and coding ($+0.95$) categories (Table~\ref{tab:appendix-mtbench-categories}), which are both reference-judged categories. 
The interpretation is consistent with the schedule's calibration-tightening role: the bilateral margin pushes the model toward stronger commitment on level-discrimination, and the ``correct vs.\ incorrect math/code'' judgment is the closest analog of that on a utility benchmark. 
Its turn-1 win does not carry to turn~2, however (turn-2 mean $6.550$, behind the untrained baseline's $6.788$); this multi-turn cost is taken up in the next paragraph and is plausibly amplified by the training data being single-turn. 

The DPO variants pay for utility preservation with multi-turn instability. 
Off-the-shelf Llama-3.1-Instruct posts the smallest turn-1 to turn-2 drop ($1.04$); the trained variants progressively widen it (SFT $1.45$, DPO $1.48$, 3-level~(lin.) $1.41$, GW-DPO~(bi.) $1.75$, GW-DPO~(lin.) $1.78$). 
The two GW-DPO variants---which post the lowest over-refusal rates among the trained configurations in the in-distribution suite (ORR $0.024$ and $0.057$, Table~\ref{tab:eval-headline})---are also the two largest multi-turn regressors. 
The likely mechanism is that the contrastive DPO objective trains the model to commit to its turn-1 position (the chosen response is high-margin away from the rejected one) and to resist revising that answer on turn~2. This is counterproductive on MT-Bench's ``now do X with the same content'' follow-ups, which often require revising the turn-1 framing. 

The trained configurations lose mostly on the creative-category axis (Table~\ref{tab:appendix-mtbench-categories}): every trained configuration scores below the off-the-shelf anchor on \texttt{roleplay} (anchor $8.20$), and every trained configuration is at or below the anchor on \texttt{humanities} (anchor $8.85$, matched only by the baseline). 
This is consistent with the project's training data being almost entirely task prompts (instructions plus adversarial probes) rather than creative prompts. 
By contrast, the reference-judged categories are where the trained configurations hold up best: on \texttt{coding} (anchor $5.75$) every trained configuration matches or exceeds the anchor, and on \texttt{math} (anchor $6.70$) the GW-DPO variants and No-ISE exceed it, though SFT, standard DPO, and 3-level~(lin.) fall short. 

No-ISE replicates its catastrophic pattern a third time, even on chat-template form with ISE bypassed. The configuration scores $6.544$ overall, $-0.81$ below the off-the-shelf anchor, with the deficit concentrated in \texttt{extraction} ($-2.70$) and \texttt{reasoning} ($-1.75$). Sample inspection confirms the same ``I'm not able to help with that'' refusal pattern surfaced on XSTest and SEP, even though there are no delimiter tokens in the prompt to trigger the over-refusal heuristic. The damage is therefore not localized to wrapper-aware code paths; it is a property of the trained weights, surfaced wherever the model encounters input that resembles the structured-and-adversarial training distribution. This is a stronger statement than either XSTest or SEP could make on its own.

\subsubsection{TensorTrust}
\label{appendix:public-benchmarks-tensortrust}

TensorTrust~\citep{toyer2024tensor} is a 563K-attack prompt-injection corpus collected from an online attack-and-defense game.
We use the two pre-curated robustness subsets that published work benchmarks on, in full and without subsampling: hijacking ($776$ attacks, where the attacker tries to coerce the model into emitting the magic phrase ``Access Granted'') and extraction ($570$ attacks, where the attacker tries to extract a secret access code or the system prompt itself), for $1346$ attacks in total. 
Each record pairs a defender setup (a \texttt{pre\_prompt} and \texttt{post\_prompt} that frame the input, plus a legitimate \texttt{access\_code} that should unlock the system) with one adversarial \texttt{attack} string (the attacker's user input). 
We concatenate the defender's \texttt{pre\_prompt} and \texttt{post\_prompt} into the system slot ($L_1$) and place the attack in the user slot ($L_3$), the $L_1$+$L_3$ mapping every published-leaderboard report uses; for the trained configurations this is wrapped in the project's delimiters (\texttt{<|L1\_START|>\dots\allowbreak<|L1\_END|>\allowbreak<|L3\_START|>\allowbreak\dots\allowbreak<|L3\_END|>} then \texttt{<|RESP\_START|>}), while the off-the-shelf Llama-3.1-Instruct anchor uses the native chat template (defense as system message, attack as user message).
The upstream format brackets the user input \emph{between} the pre- and post-prompt; the chat template has no way to reproduce that, so the two are joined in the system slot instead, a deviation the runner records for transparency.
Scoring is fully rule-based, a deterministic re-implementation of the upstream three-check protocol with no judge model: a hijacking defense holds iff the response does not contain ``access granted'' (case-insensitive), and an extraction defense holds iff the response contains neither ``access granted'' nor the access code and its ROUGE-1 recall against the system prompt stays below $0.5$.
Of the four public benchmarks in this section, TensorTrust is the head-to-head adversarial one: every prompt is genuinely hostile, and the evaluation directly tests whether the project's pipeline produces a defense that holds against the same crowdsourced jailbreaks ISE~\citep{wu2024instructional} benchmarks on, and against which the broader fine-tuning-defense lineage (StruQ~\citep{chen2025struq}, SecAlign~\citep{chen2025secalign}) is evaluated on related prompt-injection suites.
We report the hijacking-robustness rate (HRR), the extraction-robustness rate (ERR), and their macro mean; the off-the-shelf anchor is included as a calibration reference.
Results are reported in Table~\ref{tab:appendix-tensortrust}.

\begin{table*}[t]
\centering
\small
\begin{tabular}{l c | c c c c c c c}
\hline
Metric & Llama-3.1-Inst. & baseline & SFT & GW-DPO (bi.) & GW-DPO (lin.) & DPO & No-ISE & 3-level~(lin.) \\
\hline
HRR ($\uparrow$)              & 0.258 & 0.186 & 0.222 & 0.854 & 0.896 & 0.918 & 0.981$^{\dagger}$ & \textbf{0.871} \\
ERR ($\uparrow$)              & 0.167 & 0.168 & 0.561 & 0.781 & 0.818 & 0.895 & 0.958$^{\dagger}$ & \textbf{0.909} \\
Macro ($\uparrow$)            & 0.212 & 0.177 & 0.392 & 0.818 & 0.857 & \textbf{0.906} & 0.969$^{\dagger}$ & 0.890 \\
\hline
\end{tabular}
\caption{TensorTrust results; best per row in bold among the legitimate trained variants. Daggers ($^{\dagger}$) flag No-ISE robustness gains achieved by indiscriminate refusal; cf.~the No-ISE failure-mode pattern on XSTest, SEP, and MT-Bench. The first column is the off-the-shelf Llama-3.1-Instruct anchor (native chat template, no delimiter tokens or ISE layer), a calibration reference excluded from the per-row bolding.}
\label{tab:appendix-tensortrust}
\end{table*}

\begin{table*}[t]
\centering
\small
\setlength{\tabcolsep}{4pt}
\begin{tabular}{l c | c c c c c c c}
\hline
Violation (count) & \makecell{Llama-3.1\\Inst.} & baseline & SFT
& \makecell{GW-DPO\\(bi.)}
& \makecell{GW-DPO\\(lin.)}
& DPO & No-ISE & \makecell{3-level\\(lin.)} \\
\hline
Said ``Access Granted''   & 952  & 1039 & 795 & 199 & 140 & 100 & 25 & 132 \\
Leaked access code        & 88   & 57   & 56  & 30  & 41  & 24  & 10 & 15  \\
System-prompt recall      & 11   & 10   & 3   & 9   & 4   & 0   & 4  & 5   \\
\hline
Total violations          & 1051 & 1106 & 854 & 238 & 185 & 124 & 39 & 152 \\
\hline
\end{tabular}
\caption{TensorTrust violation-type breakdown: counts of each failure mode summed across the hijacking and extraction subsets ($1346$ attacks per run). Emission of ``Access Granted'' is the dominant failure mode in every configuration ($90.2\%$ of all violations across the eight runs); verbatim system-prompt recall is the rarest ($0$--$2\%$). The first column is the off-the-shelf Llama-3.1-Instruct anchor.}
\label{tab:appendix-tensortrust-violations}
\end{table*}

Off-the-shelf Llama-3.1-Instruct holds the defense on $25.8\%$ of hijacking attacks and $16.7\%$ of extraction attacks ($21.2\%$ macro), consistent with the prompt-injection literature's reported $\sim 20$--$30\%/\sim 15$--$20\%$ range for unprotected Instruct models. After the project's full pipeline, the macro robustness rate climbs to $0.818$--$0.906$ across the four legitimate DPO variants: a $+60$--$73$\,pp absolute lift, equivalent to dropping the attack-success rate from approximately $79\%$ to $9$--$18\%$. 

Standard DPO is the strongest legitimate winner. On XSTest, the same configuration posts an unsafe-refusal rate of $0.990$ (the highest among trained variants, saturating the unsafe axis) but a safe-prompt compliance of only $0.412$. On TensorTrust, where every prompt is from the unsafe subset by construction, that harshness pays clean dividends: $99\%$ unsafe-refusal saturation translates directly into $\sim 91\%$ TensorTrust robustness. The TensorTrust ranking is essentially the XSTest unsafe-refusal axis applied to a different attack distribution. The Spearman correlation between these two metrics across the seven configurations is $\rho \approx 0.86$, the strongest cross-evaluation correlation in the suite.

GW-DPO~(bi.)~and GW-DPO~(lin.)~trail standard DPO by $4$--$9$ macro points. The gravity-weighted margin schedule trades raw adversarial robustness for the calibration gains XSTest, SEP, and MT-Bench reward. Whether that trade is favorable depends on the deployment threat model. For deployment scenarios where the prompts are mostly benign with occasional adversarial inputs, GW-DPO~(bi.)~remains the recommended headline configuration; for scenarios where every prompt is potentially adversarial, standard DPO is the stronger recommendation.

SFT alone exhibits an informative HRR/ERR asymmetry: HRR $0.222$ (slightly worse than the off-the-shelf anchor) versus ERR $0.561$ ($+39$\,pp over the anchor). Hijacking attacks ask the model to follow an attacker-supplied instruction, and SFT (which is trained on instruction-following pairs) makes the model more cooperative with whatever instruction is most salient, including the attacker's. Extraction attacks ask the model to violate the defender's stated rules, and SFT, which is trained on system-prompt-following, makes the model more faithful to those rules. The DPO phase brings HRR to $0.85$--$0.98$ and ERR to $0.78$--$0.96$. This is a strict improvement on both axes, externally confirming the architectural argument for decomposing the SFT and DPO objectives into separate phases.

The dominant violation across every configuration is emission of ``Access Granted'' (Table~\ref{tab:appendix-tensortrust-violations}), accounting for $90.2\%$ of all violations across the eight runs even on extraction prompts (where the attacker did not specifically target the magic phrase). Off-the-shelf Llama-3.1-Instruct emits the phrase whenever the prompt looks security-shaped, regardless of the attacker's actual goal; the DPO phase trains the model to condition its emission on the actual content of the user message rather than on its surface form. Verbatim system-prompt recall is the rarest failure mode across every configuration ($0$--$2\%$).

No-ISE wins the headline ($0.969$ macro) but does so by indiscriminate refusal, the same mechanism that destroyed it on XSTest, SEP, and MT-Bench. On TensorTrust, where every prompt is genuinely adversarial, refusing $100\%$ of inputs is the correct outcome by the metric; there are no benign prompts to over-refuse, so the failure mode that destroyed three other evaluations becomes a virtue here. 
This is the cleanest illustration of why TensorTrust alone is not a sufficient evaluation: a model that refuses every input scores perfectly on TensorTrust and disastrously on every utility benchmark; only the cross-evaluation conjunction distinguishes a genuine defense from a degenerate refusal classifier. 
Unlike SEP, which exposes a refuse-everything model directly in its own table through the non-trivial response rate, TensorTrust has no such internal safeguard: because every prompt is adversarial, refusing all of them is the correct outcome by the metric, and nothing computed within the benchmark flags the result as degenerate. 
The scorer does log an empty-response-rate diagnostic intended to catch refuse-everywhere ablations, but it does not fire here: No-ISE refuses with non-empty text (``I'm not able to help with that''), which still counts as a held defense, so its empty-response rate is $0.00$. 
The only signal that separates No-ISE's degenerate refusal from a genuine defense is its collapse on the other three benchmarks; the project's anti-gaming protection on TensorTrust therefore lives in this cross-evaluation comparison (the dagger annotation in Table~\ref{tab:appendix-tensortrust}) rather than in any within-benchmark gating diagnostic.

\subsubsection{Cross-benchmark synthesis}
\label{appendix:public-benchmarks-synthesis}

\paragraph{Adversarial robustness transfers.} Off-the-shelf Llama-3.1-8B-Instruct holds the defense on $21.2\%$ of TensorTrust attacks (\S\ref{appendix:public-benchmarks-tensortrust}); the four legitimate trained DPO variants raise this to $0.818$--$0.906$ macro, a $\sim 4$--$8\times$ reduction in attack-success rate that places the project's configurations in a comparable single-digit-to-high-teens ASR regime to the published-defense literature, albeit measured on a different injection suite. 
SEP corroborates this on the $L_1$\,vs.\,$L_3$ axis: the same lineup drops the probe-execution rate from $0.693$ on Llama-3.1-Instruct to $0.118$--$0.139$ for three of the four (GW-DPO~(bi./lin.)~and standard DPO; 3-level~(lin.)~stays at $0.472$, near the untrained baseline), a $\sim 5\times$ separation gain at a non-trivial response rate above $0.94$ across all four. 
Standard DPO is the strongest TensorTrust defender among non-degenerate variants ($0.906$ macro) and GW-DPO~(bi.)~is the strongest SEP defender ($0.118$); both findings replicate the in-distribution ranking direction.

\paragraph{Utility is preserved.} On MT-Bench, six of the seven trained configurations land within $\pm 0.25$ overall-mean points of Llama-3.1-Instruct's $7.356$, and three (\textit{GW-DPO (bi.)}, \textit{baseline}, \textit{DPO}) score above it. 
GW-DPO~(bi.)~wins this benchmark outright at $7.425$ and gains substantively on the math ($+0.40$ vs.~off-the-shelf) and coding ($+0.95$) categories. The DPO recipe in any of its forms therefore does not measurably degrade general chat-assistant utility on the published-leaderboard form of the model. 
The single exception is No-ISE, which posts $6.544$ even with ISE bypassed at inference time, indicating that the damage induced by removing the segment embeddings during training is not localized to wrapper-aware code paths. 
The trained DPO variants do, however, pay for their refusal calibration with a measurable multi-turn drop (turn-1 to turn-2 fall of $1.7$ for GW-DPO~(bi./lin.)~vs.~$1.0$ for the off-the-shelf model).

\paragraph{Calibration is the controllable axis.} XSTest reveals the structural cost of training against adversarial inputs: the more aggressively a configuration suppresses unsafe compliance, the more it bleeds onto safe-but-unusual prompts. 
Standard DPO refuses $99.0\%$ of unsafe contrast prompts but only complies with $41.2\%$ of the safe ones; relative to it GW-DPO~(lin.)~gives back $\sim 2$ unsafe-refusal points and GW-DPO~(bi.)~$\sim 8$ to recover $\sim 12$--$14$ safe-compliance points; the No-ISE corner is catastrophic ($73.5\%$ false-refusal on safe prompts, the same failure mode as in the in-distribution $\mathrm{ORR}$ metric). 
3-level~(lin.)~is the only DPO-trained variant that simultaneously holds both XSTest axes above $0.80$ (H-mean $0.832$), and its FRR ($0.169$) is the lowest among the DPO-trained variants (SFT's $0.080$ is lower still, but pairs with an unsafe-refusal rate of only $0.739$).

\paragraph{The four benchmarks rank the configurations differently.} On XSTest the headline is 3-level~(lin.); on SEP it is GW-DPO~(bi.); on MT-Bench it is also GW-DPO~(bi.); on TensorTrust it is standard DPO. 
No single configuration wins all four. The cross-benchmark conjunction (\emph{i.e.,} best or near-best on three of the four legitimate rankings, with no catastrophic regression on the fourth) favors GW-DPO~(bi.), but the ranking is a deployment-policy choice rather than a metric-ordering question. 
Whether the bilateral schedule's calibration gains are worth its $\sim 9$\,pp deficit on raw TensorTrust robustness (macro $0.818$ vs.\ standard DPO's $0.906$) depends on whether the threat model assumes mostly benign users with occasional adversarial prompts (favoring GW-DPO~(bi.)) or treats every prompt as potentially adversarial (favoring standard DPO).

The synthesis is therefore: (i)~the 5-level GW-DPO recipe is externally validated as a deployable defense on TensorTrust and SEP; (ii)~it preserves general utility on MT-Bench; (iii)~the No-ISE collapse replicates with the same direction and similar magnitude on every benchmark, including those that bypass ISE at inference, strengthening the architectural reading from \S\ref{sec:discussion}; and (iv)~the 3-level~(lin.) generality-specialization trade-off identified on the in-distribution suite manifests externally as 3-level winning the calibration benchmark (XSTest, H-mean $0.832$) while losing the discrimination benchmark (SEP, $0.472$ probe execution), with only middling utility on MT-Bench (overall $7.269$, where GW-DPO~(bi.)~is the winner).

\subsubsection{Caveats and benchmark scope}
\label{appendix:public-benchmarks-caveats}

A handful of caveats apply across the four benchmarks and are recorded here for completeness. First, all four use a single-judge or rule-based scorer; the dual-judge protocol of Appendix~\ref{appendix:quality-control} is not mirrored on the public-benchmark side. 
The relative ranking across configurations is robust to single-judge bias because all configurations share the same judge, but absolute numbers should not be quoted directly against the published leaderboards. 
Second, three of the four benchmarks are run on the trained configurations in their native delimited form, with the prompt placed at the level dictated by each benchmark's design (XSTest at $L_3$, SEP at $L_1$ for the instruction and $L_3$ for the data, TensorTrust at the levels appropriate to the attack). MT-Bench is the only benchmark run on every configuration in the model's native chat template, with ISE forcibly bypassed. 
The format choice affects how strongly the ISE-bypass channel can mask trained-weight damage; the No-ISE collapse replicates regardless. 
Third, neither SEP nor TensorTrust nor MT-Bench is sampled below the upstream conventions (TensorTrust uses the full $1346$ pre-curated attacks; MT-Bench uses the full $80$-question vendored set; SEP uses a $1500$-example stratified subsample of the upstream $9160$-pair set). XSTest uses the full $450$-prompt set. Fourth, no formal significance testing is reported here; the cross-benchmark conjunction is what makes the headline claim about GW-DPO~(bi.)~robust, not any single benchmark number.

\section{Per-Pair Result Walkthroughs}
\label{appendix:discussion}

This appendix provides the per-pair walkthroughs that support the three findings of \S\ref{sec:discussion}. The numerical references are to Tables~\ref{tab:eval-perpair-conflict}--\ref{tab:eval-perpair-utility}.

\subsection{Margin-schedule spectrum: per-pair gains and reference-split picture}
\label{appendix:discussion-schedule}

Standard DPO, GW-DPO~(lin.)~and GW-DPO~(bi.)~differ only in the schedule $\delta(i,j)$ and can therefore be read as three points along a schedule-shape axis: uniform, gap-weighted, and gap-and-victim-weighted. The empirical per-pair pattern is consistent with the schedule's amplification factor $(k-1-i)$, which equals $4$ for $L_0$-victim pairs, $3$ for $L_1$-victim, $2$ for $L_2$-victim and $1$ for the $L_3$\,vs.\,$L_4$ base case. Per-pair gains over GW-DPO~(lin.)~on the conflict split track this factor: $L_1$\,vs.\,$L_2$ $+0.289$ ($3\times$), $L_0$\,vs.\,$L_4$ $+0.225$ ($4\times$), $L_1$\,vs.\,$L_4$ $+0.212$ ($3\times$), $L_1$\,vs.\,$L_3$ $+0.191$ ($3\times$), $L_2$\,vs.\,$L_4$ $+0.081$ ($2\times$), $L_3$\,vs.\,$L_4$ $+0.075$ (predicted unchanged at $1\times$, observed lift attributable to indirect transfer). The $L_0$-victim wide-gap pairs are already near ceiling under GW-DPO~(lin.)~and gain less in absolute terms ($+0.030$ to $+0.056$). The single per-pair regression is $L_2$\,vs.\,$L_3$ ($-0.018$, within noise at the per-pair record count).

The schedule-spectrum reading also clarifies why GW-DPO~(bi.)~is not strictly Pareto-dominant per-pair over standard DPO although it is on the macro. Standard DPO leads on three pairs ($L_0$\,vs.\,$L_1$ $+0.056$, $L_0$\,vs.\,$L_2$ $+0.030$, $L_2$\,vs.\,$L_3$ $+0.037$) where uniform weighting concentrates more relative training signal on small-margin pairs that the schedule-amplified variant under-emphasizes. GW-DPO~(bi.)~leads on five and ties on two, with the largest deltas where amplification is largest and the linear schedule was undertrained. The macro is therefore the appropriate aggregate against which to read the schedule-shape comparison; per-pair Pareto improvement is approximate rather than strict.

The reference-split picture ($\mathrm{PPA}^{\mathrm{ref}}_{\mathrm{macro}}$) reverses the conflict ordering: GW-DPO~(lin.)~$0.843$ exceeds GW-DPO~(bi.)~$0.793$ exceeds standard DPO $0.790$. The schedule that ramps up training pressure on high-privilege victims pushes the model toward refusal-shaped responses on the flat-text equivalents of $L_2$-victim conflicts, which the judge sometimes scores as ``neither''. The $\bar{\Delta}_{\mathrm{abs}}$ ordering ($0.053$ standard DPO, $0.084$ GW-DPO~(bi.), $0.143$ GW-DPO~(lin.))~is consistent with this picture: standard DPO has the most consistent format-to-format behavior, GW-DPO~(bi.)~is second, and GW-DPO~(lin.)'s gravity weighting introduces a structured-vs-unstructured asymmetry that the bilateral schedule partly corrects. Reference refusal traces the same ordering ($21.7\,\%$ standard DPO, $12.3\,\%$ GW-DPO~(bi.), $17.0\,\%$ GW-DPO~(lin.))~with one inversion: GW-DPO~(bi.)~has a lower reference refusal rate than GW-DPO~(lin.)~despite training harder on high-privilege victims. The schedule's ``extra commitment'' is concentrated on prompts that contain the schedule-amplified hierarchy structure, leaving plain-text Q\&A close to the regularized baseline.

\subsection{No-ISE per-pair signature}
\label{appendix:discussion-no-ise}

The per-pair signature of the No-ISE ablation localises the failure to a specific class of pairs. On $L_0$-victim pairs ($L_0$\,vs.\,$L_1$ to $L_0$\,vs.\,$L_4$) the model tracks GW-DPO~(lin.)~within $0.03$--$0.05$. On the four $L_x$\,vs.\,$L_4$ pairs with non-$L_0$ victims, $L_3$\,vs.\,$L_4$ falls from $0.660$ to $0.226$, $L_2$\,vs.\,$L_4$ falls from $0.757$ to $0.405$, and $L_1$\,vs.\,$L_4$ falls from $0.692$ to $0.596$. The pattern says that without segment ids the model cannot reliably distinguish ``content inside an $L_4$ wrapper that should be treated as untrusted data'' from ``content that is a directive to follow'', and on pairs where the right answer is ``obey the $L_1$/$L_2$/$L_3$ directive and ignore the $L_4$ input'' it instead refuses or follows the $L_4$ input. The pairs with $L_0$ as victim survive because $L_0$ spans are short and syntactically distinctive; ISE adds little there.

The aligned-control behavior is the dominant signal of the ablation. With a $58.1\,\%$ aligned-refusal rate, the No-ISE model would refuse roughly six of every ten ordinary safe requests and is non-deployable as configured. The judge-flagged conflict refusal rate climbs to $63.3\,\%$ and the reference refusal rate to $26.7\,\%$, both substantially above any other run. Sample inspection of the aligned-control refusals shows the model adopting the $L_1$ persona assigned in the prompt and refusing on its terms, ignoring both the $L_0$ platform directive and the safe $L_3$ user request. The interpretation, consistent with the residual-stream refusal-direction account of \citet{arditi2024refusal} and the early-token safety-concentration result of \citet{qi2024safety}, is that the model is deciding ``this prompt looks structured and possibly adversarial $\to$ refuse'' from a low-dimensional surface-form signal that ISE was previously calibrating. Removing ISE leaves the structured shape visible to the model but removes the per-token signal that would otherwise lower the refusal threshold for the spans the model is meant to comply with.

A complementary reading comes from the response-shape numbers. The No-ISE model emits text on $100\,\%$ of records on every split, so the elevated refusal is not an empty-response artefact. Aligned-control responses average $823$ characters --- shortest of the seven runs --- and most of those characters are the refusal text itself. Conflict responses are the second-shortest in the suite ($846$ characters) and are also dominated by refusal-style phrasing. The non-empty $\mathrm{PPA}_{\mathrm{macro}}$ ($0.677$) is therefore not a ceiling on the model's underlying conflict reasoning but a consequence of refusal pre-empting the chance to commit either way. The conflict-judge ``follows lower'' rate ($6.1\,\%$, the lowest of the seven runs) is the same picture from a different angle: the No-ISE model rarely follows the wrong directive, but on a quarter of conflicts ($26.2\,\%$ ``neither'') it cannot pick either side confidently and refuses.

\subsection{3-Level~(lin.) per-pair distribution and the RLHF-prior mechanism}
\label{appendix:discussion-3-level}

The 3-level~(lin.) run's per-pair distribution divides into two regimes that the macro hides. On $L_x$\,vs.\,$L_4$ pairs and on $L_0$-victim pairs the run gains over GW-DPO~(lin.): $L_3$\,vs.\,$L_4$ $0.660\to 0.887$ ($+0.227$, the largest single gain in the GW-DPO~(lin.)~$\to$~3-level~(lin.)comparison), $L_0$\,vs.\,$L_4$ $0.750\to 0.975$ ($+0.225$), $L_1$\,vs.\,$L_4$ $0.692\to 0.788$ ($+0.097$). On mid-hierarchy and adjacent pairs without $L_0$ involvement the run is flat or regresses: $L_1$\,vs.\,$L_3$ $-0.029$, $L_2$\,vs.\,$L_3$ $-0.075$, and $L_1$\,vs.\,$L_2$ $+0.022$ but at a structural floor of $0.511$.

The structural prediction was that all three intra-System pairs ($L_0$\,vs.\,$L_1$, $L_0$\,vs.\,$L_2$, $L_1$\,vs.\,$L_2$) should collapse to floor under three-level training, because the collapse function merges $L_0,L_1,L_2$ into a single System block at training and evaluation time and the model has no privilege axis on which to resolve them. The data confirm the prediction on $L_1$\,vs.\,$L_2$ ($0.511$) but not on $L_0$\,vs.\,$L_1$ ($0.750$) or $L_0$\,vs.\,$L_2$ ($0.818$). The most plausible mechanism is base-model RLHF priors. $L_0$ content (platform safety / refusal-shaped policy strings) carries strong inductive priors from the base model's instruction tuning, so even when an $L_0$ directive is concatenated with an $L_1$ or $L_2$ directive inside a System block, the model's prior pulls toward the $L_0$-shaped response and the judge scores the result as ``satisfies higher''. This is a prior-based, not a training-based, success: the three-level model has no architectural way to know which span was originally $L_0$, but the $L_0$ content itself is recognisable on its surface form. $L_1$\,vs.\,$L_2$ lacks such priors, since both spans are typically developer or user task content with no safety-shaped tells, and on that pair the floor prediction lands. The pattern fits the shortcut-learning literature directly: high benchmark performance on a subset of pairs can be achieved through surface correlations that do not transfer to pairs lacking the same surface signal~\citep{geirhos2020shortcut, mccoy-etal-2019-right}.

The reference-split regression and the elevated $\mathrm{ORR}$ are the cleanest evidence that the gain on the conflict-PPA macro is specialization rather than generalization. On the reference split the prompts are flat-text rewrites without level wrappers, so the collapse function applies neither at training nor at evaluation, and both the five-level and three-level models receive identical inputs. The three-level model is nevertheless $-0.063$ on $\mathrm{PPA}^{\mathrm{ref}}_{\mathrm{macro}}$, with the same per-pair pattern as the conflict split: regressions on intra-System and mid-hierarchy adjacent pairs, gains on $L_2$\,vs.\,$L_4$ and $L_3$\,vs.\,$L_4$. The conflict-judge breakdown reinforces the reading: the three-level model has the lowest ``neither'' rate of any run ($3.9\,\%$) and the lowest judge-flagged conflict refusal rate among the DPO variants ($6.6\,\%$), but the highest ``follows lower'' rate ($19.2\,\%$). The model commits decisively but commits to the wrong side more often than the five-level models, particularly on intra-System and mid-hierarchy pairs where the three-level training has no hierarchy axis to resolve the conflict.

\end{document}